\documentclass{article}
\usepackage{amsmath}
\usepackage{graphicx}
\usepackage{natbib}
\usepackage{url} 
\usepackage{algorithm}
\usepackage{algpseudocode}
\usepackage{float}
\usepackage{caption}

\makeatletter
\algnewcommand{\LineComment}[1]{\Statex \hskip\ALG@thistlm
 $\triangleright$ #1}
\algnewcommand{\IndentLineComment}[1]{\Statex \hskip\ALG@tlm
 $\triangleright$ #1}
\makeatother

\begin{document}

\def\half{{\scriptstyle\frac{1}{2}}}


\title{\bf Design of Hamiltonian Monte Carlo for perfect simulation of
 general continuous distributions}
\author{George M. Leigh
 \hspace{.2cm}\\
 Fisheries Queensland, Department of Agriculture and Fisheries\\
 Brisbane Qld 4000, Australia\\
 Email george.leigh@daf.qld.gov.au, george.m.leigh@gmail.com\\
 ORCiD 0000-0003-0513-2363, LinkedIn georgeleigh\\
 and\\
 Amanda R. Northrop\\
 Fisheries Queensland, Department of Agriculture and Fisheries\\
 Nambour Qld 4560, Australia\\
 ORCiD 0000-0002-8661-2459, LinkedIn amanda-northrop-a535708}
\maketitle


\bigskip
\begin{abstract}
Hamiltonian Monte Carlo (HMC) is an efficient method of simulating smooth
 distributions and has motivated the widely used No-U-turn Sampler (NUTS) and
 software Stan.  We build on NUTS and the technique of ``unbiased sampling''
 to design HMC algorithms that produce perfect simulation of general
 continuous distributions that are amenable to HMC.  Our methods enable
 separation of Markov chain Monte Carlo convergence error from experimental
 error, and thereby provide much more powerful MCMC convergence diagnostics
 than current state-of-the-art summary statistics which confound these two
 errors.  Objective comparison of different MCMC algorithms is provided by the
 number of derivative evaluations per perfect sample point.  We demonstrate
 the methodology with applications to normal, $t$ and normal mixture
 distributions up to 100 dimensions, and a 12-dimensional Bayesian Lasso
 regression.  HMC runs effectively with a goal of 20 to 30 points per
 trajectory.  Numbers of derivative evaluations per perfect sample point range
 from 390 for a univariate normal distribution to 12,000 for a 100-dimensional
 mixture of two normal distributions with modes separated by six standard
 deviations, and 22,000 for a 100-dimensional $t$-distribution with four
 degrees of freedom.
\end{abstract}

\noindent%
\vfill

\newpage

\section{Introduction}\label{secintro}

The Monte Carlo method has become an indispensable tool in science,
 engineering and finance~\citep{kroese_why_2014}.  In its ideal form
 it provides a perfect sample of independent random objects from a
 specified distribution.  Often the practitioner accepts much less
 than this ideal, especially when the only feasible means of sample
 generation is Markov chain Monte Carlo (MCMC).  Tests for convergence
 of MCMC to the target distribution are most commonly based on the
 Gelman--Rubin statistic~\citep{gelman_inference_1992}, rely on
 large-sample asymptotic decay of test statistics and confound
 convergence error with experimental error \citep[see review
 in][]{roy_convergence_2020}.  Moreover, objects generated by MCMC
 generally exhibit autocorrelation
 \citep[sec.~6.7]{gelman_inference_2011}.

Perfect simulation overcomes the above drawbacks by generating objects that
 exactly follow the target distribution and are independent \citep[see reviews
 by][]{craiu_perfection_2011, huber_perfect_2016}.  Few problems are currently
 amenable to perfect simulation.

Hamiltonian Monte Carlo (HMC)~\citep{duane_hybrid_1987,
 neal_probabilistic_1993, neal_bayesian_1995, brooks_mcmc_2011}
 efficiently simulates smooth distributions and has motivated the
 No-U-Turn Sampler (NUTS)~\citep{hoffman_no-u-turn_2014} and software
 Stan~\citep{carpenter_stan:_2017}.

This paper presents advances in design of algorithms for HMC
 (section~\ref{secfruts}) and perfect simulation
 (Section~\ref{secnearperfect}).  We build on NUTS to design two new HMC
 algorithms: the NUTS4 algorithm which operates on groups of four trajectory
 points and, through earlier detection of U-turns, reduces the number of
 discarded points; and the Full Random Uniform Trajectory Sampler (FRUTS,
 pronounced ``fruits''), which discards at most one point at each end of a
 trajectory.  We present a new method of setting the time step, a critical HMC
 parameter, and describe the numbering of HMC trajectory points which is
 integral to perfect simulation.  Sections~\ref{secapplstandard}
 and~\ref{seclasso} demonstrate the methods on a range of standard
 distributions and a challenging real-world problem.



Compared to current MCMC practice, our methods offer the advantages of
 decoupling convergence error from experimental error, replacing asymptotic
 convergence tests by perfect simulation in an exact number of iterations, and
 producing independent samples.  Our methods are widely applicable to
 continuous distributions with differentiable likelihoods.

\section{Design of Hamiltonian Monte Carlo algorithms}\label{secfruts}

\subsection{Review of Hamiltonian Monte Carlo}\label{secreviewhmc}

HMC, originally known as ``Hybrid Monte Carlo''~\citep{duane_hybrid_1987}
 from its alternation of random and deterministic steps, simulates a
 $d$-dimensional vector $q$ by introducing a time variable $t$ and a
 $d$-dimensional ``momentum'' vector $p$.  It defines the Hamiltonian as $H
 = U + K$, where potential energy $U$ is the negative log-likelihood of
 $q$, and kinetic energy $K$ is a positive, differentiable function of $p$
 satisfying $K(-p) = K(p)$.  We assume the form
\begin{equation}
 K = |p|^\beta/\beta \label{eqkpower}
\end{equation}
\noindent where $|p|$ is the Euclidean norm and $\beta > 0$.  The only
 commonly used setting is $\beta = 2$.

HMC (Algorithm~\ref{alghmc}) samples $p$ independently from the likelihood
 proportional to $\exp(-K)$ and calculates a deterministic trajectory,
 analogous to the orbit of a particle with position $q$ and momentum $p$,
 according to Hamilton's equations of motion \citep{hamilton_general_1834,
 hamilton_second_1835}:
\begin{equation}
 \dot q = \partial H / \partial p = dK / dp, \quad \dot p = -\partial
 H / \partial q = -dU/dq, \label{eqqdotpdot}
\end{equation}
\noindent where dots denote differentiation with respect to time.
 Under assumption~(\ref{eqkpower}),
\begin{equation}
 \dot q = \left(dK/{d|p|}\right) p/|p| = |p|^{\beta - 2} p. \label{eqqdotkpower}
\end{equation}

\begin{algorithm}[ht]
 \caption{Hamiltonian Monte Carlo for a general Trajectory function,
  defined later} \label{alghmc}
 \begin{algorithmic}[1]
  \State \textbf{global} $U, K, DU, DK, \delta t, d, d_R$ \Comment{Energy and
   derivatives, time step, dimensions}
  \Procedure{HMC}{$q_\text{st}, n_T, R$} \Comment{$R$: Random number
    matrix $n_T \times d_R$, indexed $[i_T, j]$}
   \State $q_0 \gets q_\text{st}$ \Comment{Set starting point.}
    \label{hmcq0assign}
   \For{$i_T \gets 1:n_T$} \Comment{$i_T$, $n_T$:
     trajectory count, number of trajectories} \label{hmcloopstart}
    \IndentLineComment{Sample momentum independently at the start of each
     trajectory.}
    \State \textbf{sample} $p_0 \sim e^{-K}$ \textbf{using} $R[i_T,
     j_\text{mom}]$ \Comment{$j_\text{mom}$: momentum random number index}
    \State $H_0 \gets U(q_0) + K(p_0)$ \Comment{Calculate Hamiltonian at
     origin.}
    \LineComment{Run the trajectory and select a destination point:
     may run for a fixed time}
    \LineComment{period (original HMC), or select randomly from
     trajectory points (NUTS).}
    \State $(q, p) \gets$ \Call{Trajectory}{$q_0, p_0,
     R[i_T, j_\text{traj}]$} \Comment{$j_\text{traj}$:
      Trajectory random numbers}
    \State $H \gets U(q) + K(p)$ \Comment{Calculate Hamiltonian at
     destination.}
    \State \textbf{if} $R[i_T, j_\text{MH}] \le \exp(H_0 - H)$ \textbf{then}
     $q_0 \gets q$ \Comment{$j_\text{MH}$: M--H test random
     number}  \label{hmcmhtest}
    \LineComment{Destination~$q_0$ is origin of next trajectory.  If M--H test
     fails, $q_0$ is unchanged.}
   \EndFor \label{hmcloopfinish}
   \State \textbf{return} $q_0$
  \EndProcedure
 \end{algorithmic}
\end{algorithm}


Hamilton’s equations~(\ref{eqqdotpdot}) endow trajectories with properties
 that ensure reversibility or ``detailed balance'' and establish that the
 stationary distribution of HMC is the target distribution~\citep[see,
 e.g.,][]{brooks_mcmc_2011}: 1.~Volume is preserved: a region in ($q$, $p$)
 space is mapped to a region with the same volume.  2.~Trajectories are
 reversible: negating the momentum at any point and applying the
 dynamics~(\ref{eqqdotpdot}) retraces the trajectory back to its origin.
 3.~The value of the Hamiltonian $H$ is preserved.

In practice, the dynamics~(\ref{eqqdotpdot}) have to be discretized in time
 and the {\it leapfrog} algorithm retains properties 1 and~2 \citep[see,
 e.g.,][]{duane_hybrid_1987}.  Given an origin $q_0$ and initial momentum
 vector $p_0$ at time zero, leapfrog calculates $q$ at integer multiples of
 the time step $\delta t$, and $p$ at integer-plus-a-half multiples.  It does
 not exactly preserve the value of $H$ but compensates by inserting a
 Metropolis--Hastings update, whereby if the value of $H$ increases by an
 amount $\delta H$ from the origin to the destination, the new value of $q$ is
 accepted with probability $e^{-\delta H}$.  In case of rejection the
 destination is reset to the origin~\citep{metropolis_equations_1953,
 hastings_monte_1970, duane_hybrid_1987}.

\subsection{New HMC algorithms}\label{secnewhmc}

The original HMC trajectories of \citet{duane_hybrid_1987}, which we call
 ``raw'' HMC, simply travelled for a fixed number of points along the
 trajectory.  For comparison with other algorithms, we set raw HMC to travel
 ten points forward and ten points backward from the origin, making a total of
 21 points, close to our target number of points in the advanced algorithms
 (NUTS, NUTS4 and FRUTS).  We chose the destination point randomly and
 uniformly from these 21 points, in the same way as the advanced algorithms
 do.

The No U-Turn Sampler~\citep{hoffman_no-u-turn_2014} was a major advance in
 HMC algorithm design.  Because NUTS discards up to half its eligible
 trajectory points, we designed two new algorithms, which we call NUTS4 and
 the Full Random Uniform Trajectory Sampler (FRUTS).  NUTS4 is a modification
 to NUTS to detect U-turns earlier and reduce the number of discarded points,
 while FRUTS is designed differently to discard at most one point at each end
 of a trajectory.  We present NUTS4 in detail in Algorithm~\ref{algnuts4}, as
 it performed the best in the majority of our tests.  The advantage of FRUTS
 in greatly reducing the number of discarded points was frequently outweighed
 by the orientation of NUTS and NUTS4 trajectories roughly parallel to the
 initial momentum, instead of random as in FRUTS.  FRUTS is described in
 detail in section~S1 of Supplementary Material.

In NUTS4 we make three innovations to NUTS.  Firstly, we follow the philosophy
 of the leapfrog algorithm in calculating the momentum vector $p$, as far as
 possible, at only integer-plus-a-half time points, as opposed to the integer
 time points used in NUTS.  This change delays termination of a trajectory
 until the discrete trajectory turns back on itself, instead of forcing
 termination as soon as a U-turn is detected in continuous time.  It also
 saves an evaluation of the derivative $dU/dq$ at each end of the trajectory,
 except when the trajectory's randomly chosen destination point happens to be
 an endpoint.

The second innovation in NUTS4 is more frequent testing for U-turns, to detect
 a U-turn earlier and reduce the number of discarded points.  A trajectory is
 divided into disjoint segments of four consecutive points which are the
 minimum units on which U-turn tests are conducted.  Our tests use the
 internal momentum vectors half-way between points 1 and~2, and between points
 3 and~4 of a segment.  The arc from every segment to every other segment is
 also tested for a U-turn.  This more intensive testing is facilitated by
 control of the number of points in a trajectory through our setting of the
 time step (see section~\ref{sectimestep}), which keeps the total number of
 tests manageable.

Our third innovation is to impose a minimum of 16 and a maximum of 256 on the
 number of points in a trajectory.  Again, this is facilitated by our setting
 of the time step, which operates to a goal of about 20 points per trajectory.
 The three innovations do not affect the proof of detailed balance for NUTS
 given by \citet{hoffman_no-u-turn_2014}, because all four-point segments, and
 pairs of segments, are treated equally.

\begin{algorithm}[ht]
 \caption{NUTS4 trajectory for use in Algorithm~\ref{alghmc}: Setup and
   forward flop} \label{algnuts4}
 \begin{algorithmic}[1]
  \Procedure{Trajectory}{$q_0, p_0, r$} \Comment{$r$: Vector of random
    numbers, indexed by $[j]$}
   \State $i^+ \gets i^- \gets 0$; $q_\text{traj}[0,] \gets q^+ \gets q^-
    \gets q_0$ \Comment{$q_\text{traj}[i^-:i^+,]$ stores trajectory points.}
   \State $p_\text{traj}[0,] \gets p_0$; $\dot p_0 \gets -DU(q_0)$;
    $\text{Uturn} \gets \textbf{false}$
    \Comment{$p_\text{traj}[i^-:i^+,]$ stores momentum.}
   \State $p^+ \gets p_0^+ \gets p_0 + \half \delta t\, \dot p_0$;
    $p^- \gets p_0^- \gets p_0 - \half \delta t\, \dot p_0$
    \Comment{$p_0^+, p_0^-$: Half time-step offset of $p_0$}
   \For{$i_\text{flop} \gets 1:n_\text{flop}$} \Comment{``Flop'' by
     $2^{i_\text{flop} - 1}$ points to one side, up to max length.}
    \If{$r[j_\text{flop} + i_\text{flop}] \ge 0.5$} \Comment{$j_\text{flop}$:
      Offset for flops; $\half$--$\half$ forward or backward}
     \State $i^+_\text{new} \gets i^+ + 2^{i_\text{flop} - 1}$ \Comment{New
      value of $i^+$, if flop doesn't produce a U-turn}
     \For{$i \gets (i^+ + 1):i^+_\text{new}$} \Comment{Hamiltonian dynamics:
       $p$ already done if $i = 1$}
      \State \textbf{if} $i > 1$ \{$\dot p \gets -DU(q^+)$; $p^+ \gets p^+ +
       \delta t\, \dot p$\} \Comment{$p^+$ is at time $(i - \half) \delta t$.}
      \State $\dot q \gets DK(p^+)$; $q^+ \gets q^+ + \delta t\, \dot q$
       \Comment{Unconditional; $q^+$ is at time $i\, \delta t$.}
      \If{$i - i^+ = 0$ \textbf{mod} 4} \Comment{Test for U-turn every fourth
        point.}
       \For{$i_\text{test} \gets i^-:(i - 3):4$} \Comment{Current and previous
	 segments}
	\State{$q_\text{span} \gets q^+ - q_\text{traj}[i_\text{test},]$;
	 $\text{Uturn} \gets q_\text{span} \cdot p_\text{traj}[i_\text{test} +
	 (i_\text{test} \ge 0),] < 0$}
	\State{$\quad \textbf{or } q_\text{span} \cdot p^+ < 0$; \textbf{if}
	 Uturn \textbf{break}} \Comment{Skip over $p_\text{traj}[0,]$.}
       \EndFor
      \EndIf \Comment{Next statement enforces minimum trajectory length 16
       points.}
      \State \textbf{if} Uturn \textbf{and} $i_\text{flop} > 4$ \textbf{break}
       \textbf{else} \{$q_\text{traj}[i,] \gets q^+$; $p_\text{traj}[i,]
       \gets p^+$\}
     \EndFor \Comment{Next: Confirm valid flop and handle flop to backward
      side.}
     \State \textbf{if} Uturn \textbf{and} $i_\text{flop} > 4$ \textbf{break}
      \textbf{else} $i^+ \gets i^+_\text{new}$ \Comment{\textbf{Alg. continues
      below.}}
  \algstore{nuts4sto}
 \end{algorithmic}
\end{algorithm}

\begin{algorithm}[ht]
 \ContinuedFloat
 \caption{NUTS4 trajectory: Backward flop and selection of destination point}
 \begin{algorithmic}[1]
  \algrestore{nuts4sto}
    \Else \Comment{Random flop direction is to backward side.}
     \State $i^-_\text{new} \gets i^- - 2^{i_\text{flop} - 1}$ \Comment{New
      value of $i^-$, if flop doesn't produce a U-turn}
     \For{$i \gets (i^- - 1):i^-_\text{new}:-1$} \Comment{Hamiltonian
       dynamics: $p$ done if $i = -1$}
      \State \textbf{if} $i < -1$ \{$\dot p \gets -DU(q^-)$; $p^- \gets p^- -
       \delta t\, \dot p$\} \Comment{$p^-$ is at time $(i + \half) \delta t$.}
      \State $\dot q \gets DK(p^-)$; $q^- \gets q^- - \delta t\, \dot q$;
       \Comment{Unconditional; $q^-$ is at time $i\, \delta t$.}
      \If{$i^- - i = 0$ \textbf{mod} 4} \Comment{Test for U-turn every fourth
        point.}
       \For{$i_\text{test} \gets i^+:(i + 3):-4$} \Comment{Current and
	 previous segments}
	\State{$q_\text{span} \gets q_\text{traj}[i_\text{test},] - q^-$;
	 $\text{Uturn} \gets q_\text{span} \cdot p_\text{traj}[i_\text{test} -
	 (i_\text{test} \le 0),] < 0$}
	\State{$\quad \textbf{or } q_\text{span} \cdot p^- < 0$; \textbf{if}
	 Uturn \textbf{break}} \Comment{Skip over $p_\text{traj}[0,]$.}
       \EndFor
      \EndIf \Comment{Next statement enforces minimum trajectory length 16
       points.}
      \State \textbf{if} Uturn \textbf{and} $i_\text{flop} > 4$ \textbf{break}
       \textbf{else} \{$q_\text{traj}[i,] \gets q^-$; $p_\text{traj}[i,] \gets
       p^-$\}
     \EndFor
     \State \textbf{if} Uturn \textbf{and} $i_\text{flop} > 4$ \textbf{break}
      \textbf{else} $i^- \gets i^-_\text{new}$ \Comment{Confirm valid flop.}
    \EndIf; \quad \textbf{if} Uturn \textbf{and} $i_\text{flop} = 4$
     \textbf{break} \Comment{Break if U-turn at 16 points.}
   \EndFor \Comment{End of $i_\text{flop}$ loop}
   \LineComment{Select the destination randomly and uniformly using variate
    $r[j_\text{sel},] \sim U(0, 1).$}
   \State $i_\text{dest} \gets i^- + \textbf{floor}(\{i^+ - i^- + 1\} \times
    r[j_\text{sel}])$ \Comment{$i^+ - i^- + 1 =$ no. of trajectory points.}
    \label{algdestselection}
   \State \textbf{return} $(q_\text{traj}[i_\text{dest},],
    p_\text{traj}[i_\text{dest},])$ \Comment{\textit{Code omitted:} Advance
    $p$ by $\pm \half \delta t$ if $i_\text{dest} \ne 0$.}
  \EndProcedure
 \end{algorithmic}
\end{algorithm}

\subsection{Setting the time step}\label{sectimestep}

We set the time step $\delta t$ to produce a practical number of trajectory
 points for effective sampling and an acceptable probability of passing the
 Metropolis--Hastings test (line~\ref{hmcmhtest} of Algorithm~\ref{alghmc}).
 The setting is derived in Supplementary section~S2:
\begin{equation}
 \delta t = 2h\beta^{1/\beta - 1} \alpha^{1/\alpha}\, \frac{\Gamma(d/\beta)}
 {\Gamma\big(\{d - 1\} / \beta + 1\big)} \cdot
 \frac{\Gamma\big(\{d - 1\}/\beta + d/\alpha + 1\big)}
 {\Gamma\big(\{d - 1\}/\beta + \{d - 1\}/\alpha + 1\big)}\,,
 \label{eqdeltatfinal}
\end{equation}
where $1 / h$ is the desired number of sample points and $\alpha$ depends
 on the target distribution which is assumed to have approximately unit
 variance.  Typically $h = 0.05$ for 20 points per trajectory, $\alpha = 2$
 for a short-tailed target distribution and $\alpha < 2$ for a long-tailed
 one.


\subsection{Selection of the destination point}

\begin{figure}[ht]
 \centering
 \includegraphics[width = 13cm, trim=0.5cm 0.8cm 1.2cm 1.7cm]
  {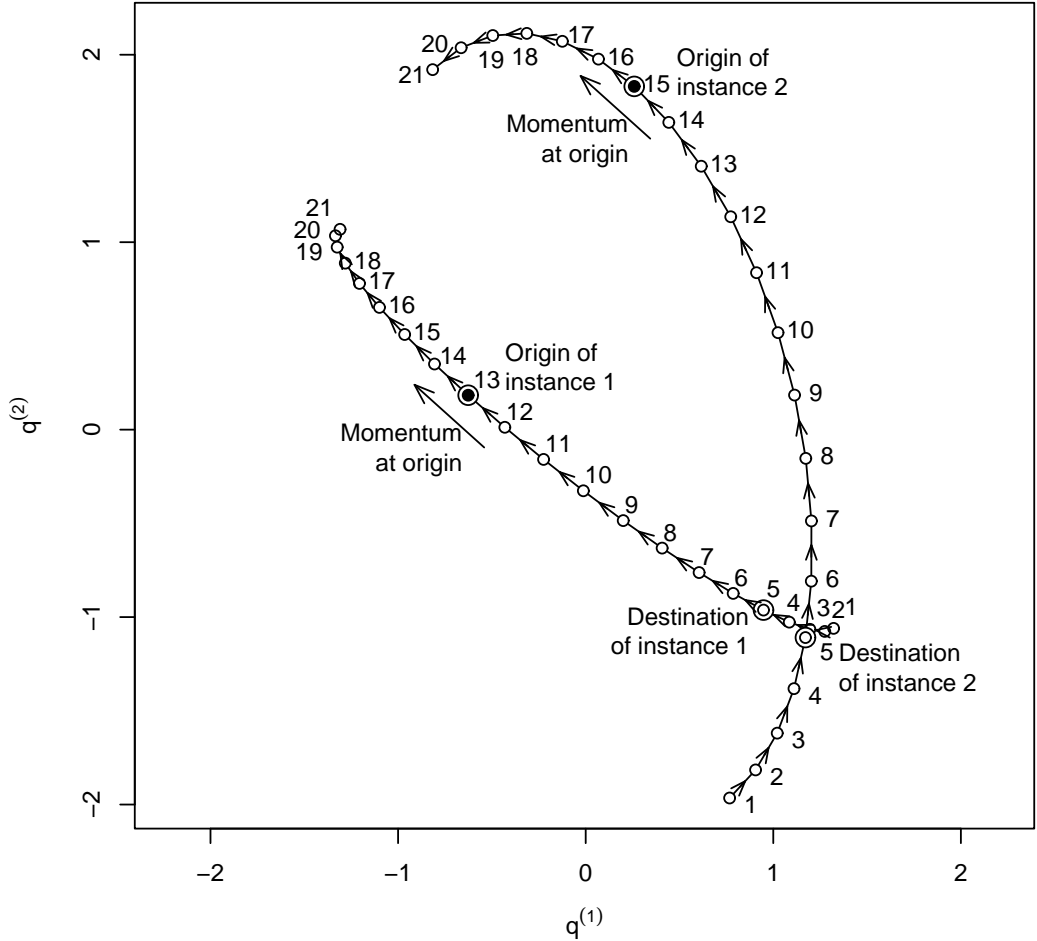}
 \caption[Design of the FRUTS algorithm]
 {Numbering of trajectory points in forward direction for coupled instances of
  a trajectory with different origins but the same random numbers.  Selection
  of the destination points is also coupled and gradually brings the instances
  closer together.}  \label{fignumberingliningup}
\end{figure}

We number the points in a trajectory in the forward direction of the
 trajectory, as shown in Figure~\ref{fignumberingliningup}.  As in NUTS, we
 select the destination randomly from the trajectory points.  We use a uniform
 random number to specify the destination point number
 (Algorithm~\ref{algnuts4}, line~\ref{algdestselection}).  Coalescence of
 coupled processes, which is enabled by this method of destination selection
 and also shown in Figure~\ref{fignumberingliningup}, will be described in
 section~\ref{seccoalescence}.

\section{Perfect simulation}\label{secnearperfect}

\subsection{Review of perfect simulation}\label{secreviewperfect}

Perfect simulation algorithms are diverse and include multigamma
 coupling~\citep{murdoch_exact_1998}, strong stationary stopping
 times~\citep{aldous_strong_1987} and perfect slice
 sampling~\citep{mira_perfect_2001}.  These algorithms are generally
 difficult to use and not widely applicable.  Multigamma coupling has an
 impossibly small probability of satisfying the conditions for perfect
 simulation for most practical problems.  Perfect slice sampling has the
 difficulty of establishing, at least approximately, the boundary of the
 feasible space, which leads to a curse of dimensionality
 \citep{bellman_review_1955} in high numbers of dimensions.

The most versatile existing perfect simulation algorithm is coupling from the
 past (CFTP) \citep{propp_exact_1996}, but even that algorithm’s applicability
 has to date been limited to a small subset of the total range of problems
 suitable for MCMC.  CFTP produces \textit{coalescence} of random objects to
 the same outcome from different starting points, and constitutes perfect
 simulation when it can be proven that the same outcome occurs from every
 possible starting point.  CFTP samples forward in time and prepends extra
 random numbers as needed to the beginning of the sequence, not the end,
 thereby avoiding bias towards ``easier'' sets of random numbers that produce
 faster coalescence.  The variant ``read-once coupling from the past''
 (ROCFTP) \citep{wilson_how_2000} proceeds only in forward sequence, in blocks
 of fixed length, some of which exhibit coalescence while others may not.

Some processes possess a \textit{monotonicity} property that allows proof of
 coalescence from every possible starting point, to establish perfect
 simulation by CFTP \citep{propp_exact_1996}.
In general, monotonicity is not available and even a discrete-valued
 process whose transition probabilities are \textit{a priori} not known
 numerically but are calculated on the run would have to visit every
 possible state to establish perfect simulation with CFTP or ROCFTP
 \citep[p.~135]{lovasz_exact_1995, fill_interruptible_1998}.  Continuous
 processes with non-countable states magnify this difficulty.

Although not presented as a perfect simulation technique, ``unbiased
 simulation'' \citep{glynn_exact_2014, jacob_unbiased_2020} estimates
 expectations of functions of coalescent random processes.  It does not
 require monotonicity or a multitude of starting points, and its absence of
 bias is exact.  Its disadvantage is that its output consists not of coalesced
 samples but of lengthy sums that can vary wildly and even produce values
 outside the allowed space, such as histogram probability values that are
 negative or greater than~1.

Unbiased simulation takes coupled random processes $X$ and $Y$ with the same
 starting point or distribution of starting points
 \citep[][p. 545]{jacob_unbiased_2020}.  The processes' random numbers are
 offset by one iteration: the transition from the starting point $Y_0$ of $Y$
 to the next value $Y_1$ uses the same random numbers as that from $X_1$ to
 $X_2$.  An unbiased sample for the expectation of the ergodic distribution of
 some function which we denote $g(X)$ is
\begin{equation}
 g(X_k) + \sum_{i = k + 1}^{\infty} \{g(X_i) - g(Y_{i - 1})\},
 \label{equnbiasedsum}
\end{equation}
where $k \ge 0$ is some chosen burn-in-like number of iterations.
 Processes $X$ and $Y$ are designed to coalesce, so that for some random
 number of iterations $\tau$, $g(X_i) = g(Y_{i - 1})$ for all $i \ge \tau$ and
 the above infinite sum actually contains only a finite number of terms.

We show in section~\ref{secunbiasedperfect} below that unbiased simulation
 becomes perfect simulation when $k$ is made large enough that the probability
 of a non-empty sum in (\ref{equnbiasedsum}) becomes vanishingly small.  We
 propose a new algorithm to take advantage of the resulting large values
 of~$k$.
 
\subsection{Coalescence in HMC}\label{seccoalescence}

Gradual coalescence over successive HMC trajectories, enabled by consistent
 numbering of trajectory points and use of the same random numbers with
 different starting points, has been illustrated in
 Figure~\ref{fignumberingliningup}.
We convert gradual coalescence into exact coalescence by a final rounding step
 to a congruence (Algorithm~\ref{algcoalhmc}).  For detailed balance, the
 desired congruence in each coordinate is sampled randomly and uniformly
 modulo some small interval width~$w$, similarly to a random, uniform
 Metropolis--Hastings jump.  Any change in negative log-likelihood~$U$ is
 accounted for by a Metropolis--Hastings test.


\begin{algorithm}[hb]
 \caption{Coalescence in HMC, calling HMC (Algorithm~\ref{alghmc}) and
   rounding}
  \label{algcoalhmc}
 \begin{algorithmic}[1]
  \Procedure{HmcRound}{$q_\text{st}, n_T, R, r_\text{ro}, w$}
    \Comment{$r_\text{ro}$, $w$: rounding random vector \& width}
   \State $q \gets \Call{HMC}{q_\text{st}, n_T, R}$
    \Comment{$q_\text{st}$, $R$: initial value, matrix of random numbers}
   \State $q_\text{ro} \gets w \times \{\textbf{floor}(q / w) +
    r_\text{ro}[1:d]\}$ \Comment{Make outcomes identical if already
    close.} \label{finalrounding}
   \State \textbf{if} $r_\text{ro}[d + 1] \le \exp\{U(q) - U(q_\text{ro})\}$
    \textbf{then} $q \gets q_\text{ro}$ \Comment{M--H test for rounding}
   \LineComment{Note $r_\text{ro}$ has length $d + 1$.  Failure of M--H test
    means no exact coalescence.} \label{finalroundingend}
   \State \textbf{return} $q$
  \EndProcedure
 \end{algorithmic}
\end{algorithm}

Challenging starting points should be chosen to explore coalescence in
 problems in which no monotonicity property is available (see
 section~\ref{secreviewperfect}).  For HMC, challenging points comprise both
 extreme points and points of high likelihood where potential energy~$U$ is
 low and may provide insufficient energy to reach a destination point that is
 reachable from an extreme starting point.  Our CFTP examples use $m =
 \min(2^d + 1, 33)$ starting points, comprising $m - 1$ extreme points and the
 maximum-likelihood point; the mixture examples add an extra
 equal-maximum-likelihood point.  Coordinate values of extreme points are $\pm
 6$ in scaled coordinates that have approximately unit variance (see
 section~\ref{sectimestep} and Supplementary section~S2).  For $d \le 5$ the
 extreme points follow a $2^d$ factorial design, while for $d > 5$ the first
 five co-ordinates are factorial and the others are set randomly to $\pm 6$.

\subsection{From unbiased simulation to perfect simulation}
 \label{secunbiasedperfect}

The sample point~(\ref{equnbiasedsum}) is valid for any function~$g$,
 including a histogram with arbitrarily narrow intervals.  Therefore the
 ``string'' of pairs of values and sample weights $\{(X_k, 1)$, $(Y_k, -1)$,
 $\ldots$, $(X_{\tau - 2}, 1)$, $(Y_{\tau - 2}, -1)$, $(X_{\tau - 1}, 1)\}$ is
 an unbiased estimator of the distribution of~$X$; i.e., it constitutes a
 perfect sample point from the distribution, with the proviso that it may
 contain multiple points and negative weights.  A sample weight of $-1$
 indicates a ``hole'' that, in a large sample of many strings, will be
 cancelled out by an equal value (or a value in the same histogram bar) that
 has a weight of $+1$ in some other string.

Negative weights do not occur if the sum in~(\ref{equnbiasedsum}) is empty;
 i.e., if the processes $X$ and~$Y$ coalesce within $k$ steps.  Therefore a
 perfect sample is obtained if $k$ is chosen large enough that the probability
 of lack of coalescence within $k$ steps is vanishingly small; e.g, smaller
 than the numerical precision used in the arithmetic.

We emphasize that a target distribution should be thoroughly investigated both
 theoretically and numerically before the value of $k$ is set.  The
 consequences of negative weights are severe, as a single string can be very
 long and contain many negative weights, e.g., when one region of the sample
 space is almost cut off from the rest in a multi-modal distribution, and this
 is not appreciated before setting~$k$.  Negative weights express the reality
 that, when theory such as monotonicity from a finite set of starting points
 is not available, perfect simulation requires exploration of the entire
 sample space (see section~\ref{secreviewperfect}).  The potential for
 negative weights provides reassurance that unbiased sampling operates within
 established laws and can produce perfect simulation without visiting every
 possible state of a process.

Algorithm~\ref{algcoalhmc} substitutes an ``iteration'' in
 (\ref{equnbiasedsum}) by a ``block'' of $n_T$ HMC trajectories, followed by a
 single rounding step.  Our examples set $n_T$ so that approximately 90\% of
 combinations of our CFTP starting points and random numbers coalesce in one
 block; the remaining 10\% of combinations require more than one block.
 
\begin{figure}[ht]
 \centering
 \includegraphics[width=5.25cm, trim=2.1cm 7.5cm 7cm 1.5cm,
  clip=true]{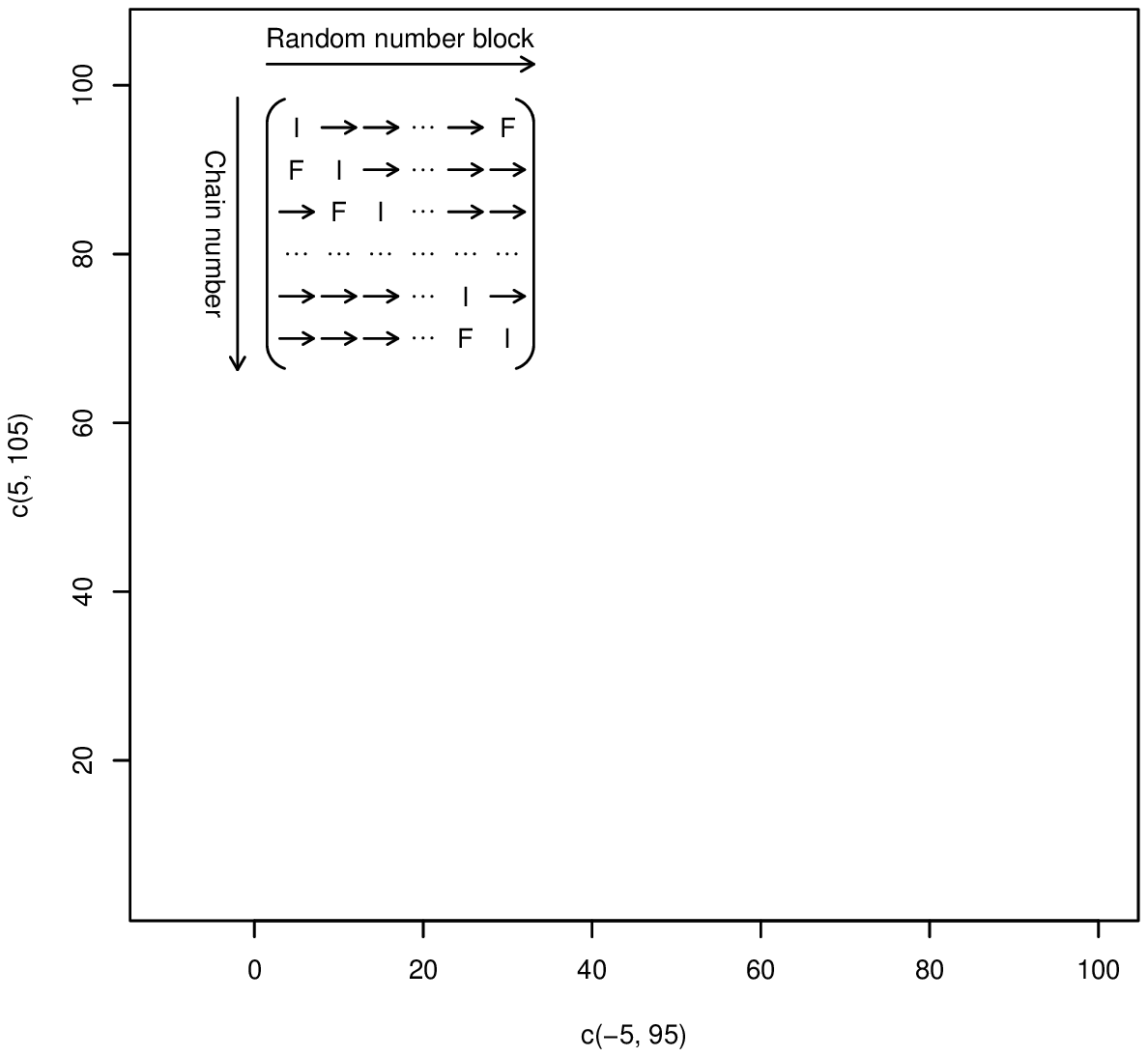}
 \caption[Chain $\times$ block matrix design]
 {Chain $\times$ block matrix for one sample set.  Random-number blocks
  proceed from left to right and are common to all chains.  Each chain is
  initialized on its diagonal element ``I'' to a starting point from some
  preset distribution.  Blocks wrap around when necessary.  The final point
  ``F'' is a perfect sample if the chain has coalesced with the one below it.}
  \label{figsampleset}
\end{figure}

Figure~\ref{figsampleset} and Algorithm~\ref{algunbiased} show how to produce
 a set of $k$ perfect sample points with only slightly more computation than
 for a single one.  Our examples take $k = 14$.  Most chains coalesce in one
 block, so that only the first and the freshly-initialized chains need to be
 run; the other chains are copies of the first.  The $k$ sample points within
 the set are not completely independent but are well separated using the block
 length as a thinning ratio.

\begin{algorithm}[!ht]
 \caption{Perfect sampling from unbiased sampling: Chain $\times$ block upper
  triangle} \label{algunbiased}
 \begin{algorithmic}[1]
  \Procedure{Unbiased}{$n_S, n_B, n_T, w$} \Comment{$n_S, n_B, n_T$:
    numbers of independent}
   \LineComment{sample sets, blocks per sample set, trajectories per block;
    $w$: rounding width}
   \State \Call{Error}{} $\gets$ \Call{False}{} \Comment{Initialize record of
    coalescence errors (occur if $n_B$ is too small).}
   \For{$i_S \gets 1:n_S$} \Comment{$i_S:$ sample set, each with its own
     chain $\times$ block matrix}
    \State $Q_0 \gets$ \Call{SampleStart}{$n_B$} \Comment{$Q_0$: Sample
     starting points for chains, $n_B \times d$}
    \State $s \gets$ \Call{RandomSeed}{}; $c \gets \textbf{zeros}(n_B)$
     \Comment{Store seed and mark all chains as active.}
    \For{$i_B \gets 1:n_B$} \Comment{Upper triangle of matrix; $i_B$: block
     number}
     \State $R \gets$ \Call{SampleHmc}{$n_T$} \Comment{Generate random numbers
      for HMC ($n_T \times d_R$).}
     \State $r_\text{ro} \gets$ \Call{SampleRound}{} \Comment{Generate random
      numbers for rounding ($d + 1$).}
     \State $Q[i_B,] \gets Q_0[i_B,]$ \Comment{Initialize chain~$i_B$
      only when we reach block~$i_B$.}
     \For{$i_C \gets 1:i_B$} \Comment{Loop over current chains; $i_C$: chain
       no.}
      \State \textbf{if} {$c[i_C] > 0$ \textbf{then} $Q[i_C,] \gets
       Q[c[i_C],]$ \textbf{else}} \Comment{Copy if inactive (coalesced).}
       \State \hskip 1.5em $Q[i_C,] \gets$ \Call{HmcRound}{$Q[i_C,], n_T, R,
        r_\text{ro}, w$} \Comment{Run HMC (Alg.~\ref{algcoalhmc}).}
       \State \hskip 1.5em \textbf{for} {$j_C \gets 1:(i_C - 1)$}
        \Comment{Record if $i_C$ has newly coalesced to chain~$j_C$.}
       \State \hskip 3em \textbf{if} $c[j_C] = 0$ \textbf{and} $Q[i_C,] =
        Q[j_C,]$ \textbf{then} $c[i_C] \gets j_C$; \textbf{break}
       \State \hskip 1.5em \textbf{end for}
      \State \textbf{end if}
     \EndFor
     \State $Q_0[i_B,] \gets Q[1,]$ \Comment{Save chain~1 for use in lower
      triangle.  Reuse array~$Q_0$.}
    \EndFor \Comment{$i_B$}
    \State $Q_\text{sto}[(iS - 1) \times n_B + 1,] = Q[1,]$ \Comment{Store
     chain~1 final.  \textbf{Continued below.}
  \algstore{unbiasedsto}
 \end{algorithmic}
\end{algorithm}

\begin{algorithm}[ht]
 \ContinuedFloat
 \caption{Perfect sampling from unbiased sampling: Chain $\times$ block lower
  triangle}
 \begin{algorithmic}[1]
  \algrestore{unbiasedsto}
    \State \Call{RandomSeed}{} $\gets s$ \Comment{Restore seed so blocks'
     random numbers are the same.}
    \For{$i_B \gets 1:(n_B - 1)$} \Comment{Lower triangle of chain $\times$
      block matrix}
     \State \Call{Error}{} $\gets$ \Call{Error}{} \textbf{or} $Q[i_B,] \ne
      Q[i_B + 1,]$ \Comment{Check coalescence of chain~$i_B$.}}
     \LineComment{\textit{Non-germane code omitted:} Rework $c$ so no chain
      $i_C > i_B$ coalesces to $i_B$.}
     \State $R \gets$ \Call{SampleHmc}{$n_T$}; $r_\text{ro} \gets$
      \Call{SampleRound}{}
     \State $Q[1,] \gets Q_0[i_B,]$ \Comment{Restore the saved value of
      chain~1 for block~$i_B$.}
     \For{$i_C \gets (i_B + 1):n_B$} \Comment{Loop over current chains.}
      \State \textbf{if} {$c[i_C] > 0$ \textbf{then} $Q[i_C,] \gets
       Q[c[i_C],]$ \textbf{else}} \Comment{Copy if inactive (coalesced).}
       \State \hskip 1.5em $Q[i_C,] \gets$ \Call{HmcRound}{$Q[i_C,], n_T, R,
        r_\text{ro}, w$} \Comment{Run HMC (Alg.~\ref{algcoalhmc}).}
       \State \hskip 1.5em \textbf{for} {$j_C \gets (1, (i_B + 1):(i_C - 1))$}
        \Comment{Record if $i_C$ has newly coalesced.}
       \State \hskip 3em \textbf{if} $c[j_C] = 0$ \textbf{and} $Q[i_C,] =
        Q[j_C,]$ \textbf{then} $c[i_C] \gets j_C$; \textbf{break}
       \State \hskip 1.5em \textbf{end for}
      \State \textbf{end if} \Comment{No HMC is needed if all chains have
       coalesced to chain~1.}
     \EndFor
     \State $Q_\text{sto}[(iS - 1) \times n_B + i_B + 1,] = Q[i_B + 1,]$
      \Comment{Store chain~$i_B + 1$ final.}
    \EndFor \Comment{$i_B$}
    \State \Call{Error}{} $\gets$ \Call{Error}{} \textbf{or} $Q[n_B,] \ne
     Q[1,]$ \Comment{Check coalescence of chain~$n_B$.}
   \EndFor \Comment{$i_S$}
   \State \textbf{return} $Q_\text{sto}$
  \EndProcedure
 \end{algorithmic}
\end{algorithm}

\section{Application to standard distributions}\label{secapplstandard}

\subsection{Methods}\label{secapplstandardmethods}

Algorithms~\ref{alghmc}, \ref{algcoalhmc} and~\ref{algunbiased} for HMC,
 coalescence and unbiased sampling, with time step given
 by~(\ref{eqdeltatfinal}), were coded in the software R
 \citep{r_core_team_r_2022}.  Four different HMC trajectory algorithms were
 coded: raw HMC \citep{duane_hybrid_1987}, NUTS
 \citep{hoffman_no-u-turn_2014}, NUTS4 (Algorithm~\ref{algnuts4}) and FRUTS
 (Supplementary Algorithm~S1).

To compare algorithms, we introduce the principle of measuring, per
 perfect-sample point, the number of evaluations of the derivative of the
 negative log-likelihood function $U$.  This derivative is used in the
 Hamiltonian dynamics (\ref{eqqdotpdot}) and constitutes the major
 computational load in practical MCMC when $U$ contains more complexity than
 the other quantities used in MCMC.

Coordinates were scaled by the square-root of the Hessian matrix at the
 maximum-likelihood point (see Supplementary section~S2), except for a
 correlated-normal example with standard normal marginal distributions but
 deliberate nonzero correlations between the coordinates.  Coupling from the
 past (Supplementary Algorithm~S2) was coded for exploratory analysis prior to
 application of unbiased sampling.

We set the time step with a goal of 20 points per trajectory; i.e., $h =
 0.05$ in~(\ref{eqdeltatfinal}).  This setting allows little room for
 misspecification of the variance in scaling the coordinates.  Difficult
 cases may require $h < 0.05$ to provide enough sample points and
 sufficiently small variation in the Hamiltonian $H$ over a trajectory.  We
 set the final coordinate-rounding parameter $w$ (line~\ref{finalrounding}
 of Algorithm~\ref{algcoalhmc}) to the value 0.01 in all our examples.

Exploratory CFTP used 20 different sets of random numbers or ``runs'' crossed
 with the $m$ starting points described in section~\ref{seccoalescence}.  We
 recorded how many trajectories were needed to attain coalescence from all
 starting points in all runs, and how many were needed for coalescence of 90\%
 of combinations of run and starting point.  The latter number set the block
 size for unbiased sampling (Algorithm~\ref{algunbiased}).  Coordinates of
 starting points for unbiased sampling (diagonal cells labelled ``I'' in
 Figure~\ref{figsampleset}) were randomly set to $\pm 6$.

We used the value 2 for the HMC energy parameter~$\beta$ in
 equation~(\ref{eqkpower}).  This is the setting almost universally used in
 HMC practice.  Preliminary experiments found that values less than~2 (e.g., 1
 or 0.5), when applied early in the sequence of trajectories (e.g., the first
 25\% of trajectories), could slightly reduce the number of trajectories
 required for coalescence.  A side effect, however, was loss of control over
 the number of points in a trajectory, i.e., ineffectiveness of the time-step
 setting in achieving its goal, which meant that the overall amount of
 computation was not significantly reduced.  We did, however, make use of
 values less than 2 for the likelihood-tail parameter $\alpha$ in
 equation~(\ref{eqdeltatfinal}), for long-tailed target distributions.

The standard distributions we tested were the standard normal
 distribution, normal distribution with correlated coordinates,
 $t$-distributions with four degrees of freedom, and mixtures of two
 standard normal distributions.  All were tested up to 100 dimensions.

The correlated normal distributions had standard normal marginal
 distributions with a fixed positive correlation $\rho$ for each pair of
 coordinates.  High correlations are not likely in practice if the
 coordinates are scaled appropriately (see above and Supplementary
 section~S2), but we included them to gauge the robustness of the method to
 correlations.

The multivariate $t$-distribution was included as an example of a
 long-tailed distribution.  Its likelihood is 
 $\big[ \Gamma\big(\{\nu + d\}/2\big)
 \big/ \big\{ (\pi\nu)^{d/2} \Gamma(\nu/2) \big\} \big]
 \left(1 + q^2/\nu\right)^{-(\nu + d)/2}$
 \citep[see, e.g., ][]{sutradhar_multivariate_2006},
 where $\nu$ is the degrees of freedom parameter and $q^2$ is the
 $d$-dimensional dot product.

The multimodal normal mixture distribution was included as a challenge to MCMC
 to travel between one mode at zero and another mode at $\mu > 0$ in the first
 coordinate.  The positive extreme CFTP starting points in coordinate 1 were
 adjusted to $-6$ and $\mu + 6$ to keep them extreme.  The mixing proportions
 were 0.5 (equal proportions) in all cases.


\subsection{Results}\label{secapplstandardresults}

Perfect simulation is easily achievable for most standard distributions, up to
 at least $d = 100$ dimensions (Table~\ref{tabresultsstandard}).  For the
 standard normal with $d = 100$, coalescence with the NUTS4 HMC algorithm was
 achieved in an average of 1.03 blocks, each of 38 trajectories.  Trajectories
 contained on average 14.0 likelihood-derivative evaluations with 8.5
 evaluations for discarded points, thereby averaging 878 derivative
 evaluations per perfect sample point.  Algorithm~\ref{algunbiased} was still
 run for the full 14 blocks, to meet the conditions for perfect sampling and
 to generate results for all 14 chains in each independent sample set.

\begin{table}[th]
 \caption[Results for standard distributions]
 {Perfect simulation results for standard distributions.  Symbol $d$ is number
  of dimensions, $\rho$ is pairwise correlation coefficient (aspect ratio of
  variance ellipsoid in parentheses), $\nu$ is degrees of freedom, $\alpha$ is
  the distribution-tail parameter (noted when $\alpha \ne 2$), and $\mu$ is
  the nonzero mode.  The perfect-sample size is listed (number of independent
  sample sets in parentheses, k denotes 1000), then the numbers of derivative
  evaluations per perfect-sample point for the NUTS4 and FRUTS algorithms (the
  lower number in bold), and the maximum number of blocks taken for chains to
  coalesce in Algorithm~\ref{algunbiased}.}
  \label{tabresultsstandard}
 \vspace{0.4 cm} 
 \centering
 \begin{tabular}{l r l r r r r r}
 \hline
 Name & $d$ & Parameters set & Sample & NUTS4 & FRUTS & Max \\
 \hline
 Standard normal & 1 & -- & 140k (10k) & \textbf{388} & 391 & 4 \\
  & 10 & -- & 140k (10k) & \textbf{552} & 1114 & 4 \\
  & 100 & -- & 140k (10k) & \textbf{878} & 1342 & 4 \\
 Correlated normal
  & 2 & $\rho = 0.6$ (4) & 140k (10k) & \textbf{528} & 825 & 3 \\
  &  & $\rho = 0.95$ (39) & 140k (10k) & \textbf{687} & 1093 & 4 \\
  & 10 & $\rho = 0.1$ (2.11) & 140k (10k) & \textbf{1011} & 1417 & 3 \\
  &  & $\rho = 0.6$ (16) & 140k (10k) & \textbf{2780} & 3955 & 4 \\
  &  & $\rho = 0.95$ (191) & 140k (10k) & \textbf{4467} & 7379 & 4 \\
  & 100 & $\rho = 0.1$ (12.11) & 70k (5k) & 7546 & \textbf{4933} & 3 \\
  &  & $\rho = 0.45$ (82.82) & 28k (2k) & 27151 & \textbf{23404} & 3 \\
 $t$-distribution & 1 & $\nu = 4$ & 140k (10k) & 919 & \textbf{575} & 4 \\
  & 10 & $\nu = 4$, $\alpha = 1.5$ & 140k (10k) & \textbf{3441} & 5986 & 4 \\
  & 100 & $\nu = 4$, $\alpha = 1.25$ & 14k (1k) & \textbf{21673} & 56557 & 5 \\
 Normal mixture & 1 & $\mu = 4$ & 140k (10k) & 1194 & \textbf{769} & 5 \\
  & & $\mu = 6$ & 140k (10k) & 5592 & \textbf{2456} & 8 \\
  & 10 & $\mu = 4$ & 140k (10k) & \textbf{1664} & 2148 & 5 \\
  & & $\mu = 6$ & 140k (10k) & \textbf{9442} & 13200 & 5 \\
  & 100 & $\mu = 4$ & 140k (10k) & \textbf{1932} & 1935 & 5 \\
  & & $\mu = 6$ & 14k (1k) & 13580 & \textbf{12061} & 5 \\
 \hline
 \end{tabular}
\end{table}

The NUTS4 algorithm had the lowest (best) number of derivative evaluations in
 the majority of our examples.  FRUTS was lower in one-dimensional and some
 100-dimensional examples, so the choice between NUTS4 and FRUTS is not
 clear-cut.  NUTS had higher numbers than NUTS4, by factors of between 1.1 and
 3, in all of the examples we tested.  Raw HMC was competitive for normal
 distributions but did not achieve coalescence reliably for the $t$ and normal
 mixture distributions.

The maximum number of blocks taken to coalesce in any of our examples was 8,
 demonstrating the margin of safety in our setting of 14 blocks in
 Algorithm~\ref{algunbiased}.

Results for the correlated normal distributions show that the algorithms are
 not greatly disturbed by moderate correlations in the coordinates.  With $d$
 = 100, a correlation of 0.1 in each pair of coordinates produces a variance
 ellipsoid of aspect ratio $\{1 + (d - 1)\rho\}/(1 - \rho)$ = 12.11, but the
 number of derivative evaluations per perfect sample point, compared to zero
 correlation, increased only from 878 to 4933 (FRUTS HMC algorithm) or 7546
 (NUTS4).  A correlation of 0.45 produces an extreme aspect ratio of 82.82 but
 the numbers of evaluations remained tractable at 23404 (FRUTS) and 27151
 (NUTS4).

For the long-tailed $t$-distributions, setting $\alpha$ to 1.5 or~1.25 in the
 time step (equation~(\ref{eqdeltatfinal})) maintained a range of about 20 to
 100 points per trajectory.  This saved computational effort compared to the
 short-tail setting $\alpha = 2$, which, in 100 dimensions, generated 70 or
 more points in most trajectories.

The $t$-distributions with $\nu = 4$ were tractable up to $d = 100$, with
 21673 derivative evaluations per perfect sample point for NUTS4, and 56557
 for FRUTS.

Normal mixtures were tractable up to at least $\mu = 6$, where the intermodal
 trough rises no higher than 2\% of the modal likelihood and descends
 extremely low on all paths that deviate substantially from the straight line
 between the modes.  The case $d = 100$ and $\mu = 6$ required 12061 (FRUTS)
 or 13580 (NUTS4) derivative evaluations per perfect sample point.

\section{Application to the Bayesian Lasso}\label{seclasso}

The Lasso~\citep{tibshirani_regression_1996} was developed as an
 alternative to subset selection in regression.  Its Bayesian
 version~\citep{park_bayesian_2008} is a challenge for HMC, due to
 discontinuities in the derivatives of the likelihood.  Perfect
 simulation for the Bayesian Lasso has been discussed
 by \citet{botev_exact_2018}.  The likelihood is
\begin{equation}
 \sigma^{-1} \cdot (2\pi)^{-n/2} \sigma^{-n} \exp\big(-{\scriptstyle
 \frac 1 2} S/\sigma^2\big) \cdot \big({\scriptstyle \frac 1 2}
 \lambda/\sigma \big)^J \exp(-\lambda T/\sigma)\,,
 \label{eqlassolik}
\end{equation}
\noindent where $\sigma$ is the standard deviation, $n$ is the number
 of data records, $S$ is the residual sum of squares, $\lambda$ is
 the Lasso parameter, $J$ is the number of regression coefficients
 excluding the intercept, and $T$ is the sum of their absolute values:
 $S = \sum_{i=1}^n {(y_i - \beta_0 -X_i \beta)^2}$ and $T =
 \sum_{j=1}^J |\beta_j|$, where $y_i$ is the $i$th value of the
 response, $X_i$ is the $i$th row of the design matrix, $\beta_0$ is
 the intercept and $\beta$ is the vector of regression coefficients
 $\beta_j$ ($j = 1$, \ldots,~$J$).

The factors in~(\ref{eqlassolik}) are the uninformative prior for
 $\sigma$; the ordinary least squares (OLS) likelihood for $\beta_0$,
 $\beta$ and $\sigma$; and the Lasso likelihood for $\beta$ and
 $\sigma$ to penalise large values of $T$ by an amount specified
 by~$\lambda$.  We use $\log\sigma$ in place of $\sigma$, so omit the
 first factor $1/\sigma$.

We do not deal with the estimation of the Lasso parameter $\lambda$.
 Rather, we explore simulation with different values of $\lambda$, and
 illustrate the increasing difficulty of the problem as $\lambda$ increases
 and the discontinuities in the likelihood derivatives become more
 important.

The model contains $J + 2$ parameters, comprising $\beta_j$ ($j =
 1$,~\ldots, $J$), $\beta_0$ and $\log\sigma$.

We applied the NUTS4 and FRUTS HMC algorithms and Algorithm~\ref{algunbiased}
 to the diabetes data set of \citet{efron_least_2004}.  This data set has 442
 rows of data and 10 explanatory variables, making 12 model parameters once
 $\beta_0$ and $\log\sigma$ are included.  We scaled all the explanatory
 variables to have mean zero and variance 1, and transformed the parameter
 vector to have approximately zero mean and identity-matrix variance by
 subtracting the OLS estimates and scaling by the square-root of the OLS
 Hessian matrix.

We tested six different values of the Lasso parameter $\lambda$: 0, 0.237
 (optimal value from Park and Casella), 1, 2, 5 and 10.  We could not achieve
 coalescence with $\lambda \ge 20$, presumably due to the increased importance
 of discontinuities in the likelihood derivative.

Results are summarized in Table~\ref{tabresultslasso}.  NUTS4 always had a
 lower number of derivative evaluations than FRUTS, as it did for the
 ten-dimensional standard distributions in
 section~\ref{secapplstandardresults}.  The number of evaluations per
 perfect-sample point increased from 704 for $\lambda = 0$ to 5822 for
 $\lambda = 10$.  We could not achieve coalescence with FRUTS for $\lambda =
 10$.

\begin{table}[th]
 \caption[Coalescence results for the Bayesian Lasso]
 {Perfect simulation results for the Bayesian Lasso with the diabetes data
  from \citet{efron_least_2004}.  Columns are the value of the Lasso parameter
  $\lambda$, perfect-sample size (number of independent sample sets in
  parentheses), average number of derivative evaluations for the NUTS4 and
  FRUTS algorithms, and the maximum number of blocks taken for a chain to
  coalesce in Algorithm~\ref{algunbiased}.}
  \label{tabresultslasso}
  \vspace{0.4 cm} 
 \centering
 \begin{tabular}{l r r r r}
 \hline
 \phantom{0}$\lambda$ & Sample & NUTS4 & FRUTS & Max \\
 \hline
 \phantom{0}0 & 140k (10k) & 704 & 1205 & 4 \\
 \phantom{0}0.237 & 140k (10k) & 708 & 1209 & 3 \\
 \phantom{0}1 & 140k (10k) & 728 & 1424 & 4 \\
 \phantom{0}2 & 140k (10k) & 804 & 2265 & 4 \\
 \phantom{0}5 & 140k (10k) & 1517 & 11077 & 7 \\
 10 & 140k (10k) & 5822 & -- & 5 \\
 \hline
 \end{tabular}
\end{table}
 
\begin{figure}[ph]
 \centering
 \includegraphics[width = 13.32cm,
  trim = 0cm 0.3cm 0cm 0.6cm]{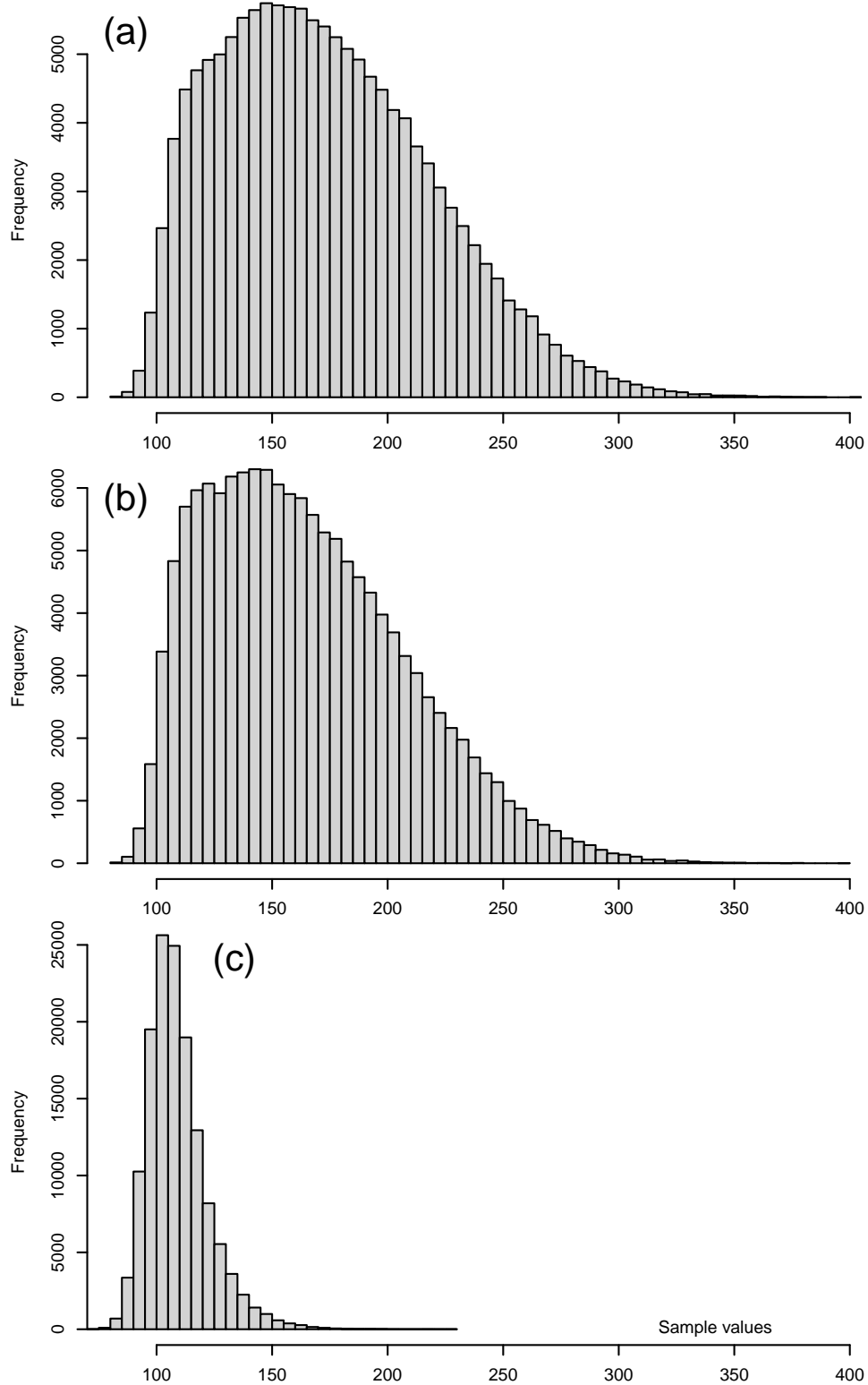}
 \caption[Histograms of sum of Bayesian Lasso regression coefficients]
 {Perfect-sample histograms of the sum of absolute values of Bayesian Lasso
  regression coefficients, $T$, for different settings of the Lasso parameter
  $\lambda$: (a)~$\lambda = 0$, (b)~$\lambda = 0.237$, (c)~$\lambda =
  5$.} \label{fighistlassot}
\end{figure}

Histograms of the sum of regression coefficients, $T$, are presented in
 Figure~\ref{fighistlassot}, for $\lambda = 0$ (no Lasso term), 0.237 and 5.
 They show little contraction in the distribution from $\lambda = 0$ to
 $\lambda = 0.237$, and a very large contraction at $\lambda = 5$.  The
 minimum value of $T$ over all the simulations moved from 82.1 at $\lambda =
 0$ to~73.6 at $\lambda = 5$, while the maximum moved from 403.3 to~228.5: the
 large value of $\lambda$ removed large values of $T$ from the distribution,
 and focussed mainly on already-existing small values.  Histograms of the
 residual sum of squares, $S$, show little change over this range of
 $\lambda$; i.e., little worsening of the fit (see Supplementary Figure~S2).

Our conclusion is that the explanatory variables in the diabetes data
 set contain appreciable redundancy, and a value of $\lambda$ that
 substantially reduces $T$ is justified.  Values of $\lambda$ greater
 than the one of 0.237 from \citet{park_bayesian_2008} may be
 appropriate.

\section{Discussion}\label{secdiscussion}

We have demonstrated perfect simulation for fairly general continuous
 distributions up to 100 dimensions.  This greatly expands the range of
 problems to which perfect simulation can be applied.  It also qualitatively
 improves MCMC convergence testing by removing the confounding between
 convergence and experimental error, which is present in all widely-used,
 existing tests of MCMC convergence.  The MCMC practitioner using existing
 tests never knows whether convergence has been achieved.  We have also
 introduced an objective measure of MCMC algorithm performance, in the number
 of likelihood derivative evaluations per perfect-sample point.
 
Our results confirm that Hamiltonian Monte Carlo is a very powerful method of
 MCMC and can be promoted over random-walk-based methods with their incumbent
 slow exploration of the sample space
 \citep[see][sec. 5.3.3]{brooks_mcmc_2011}.  HMC is still not infallible:
 setting the time step is difficult when the variance is subject to radical
 change; e.g., the ``funnel'' example of \citet[sec. 8]{neal_slice_2003-1}.
 The need for transformation and reprogramming would, however, be evident from
 lack of coalescence.

HMC practice leads theory.  Although its numerical results are outstanding,
 HMC does not produce an expression for the transition kernel, nor for the
 number of iterations that it takes to converge.

We ascribe the frequent superiority of the NUTS4 algorithm over FRUTS to its
 reliance on the initial HMC momentum vector to set the orientation of a
 trajectory.  This feature stems from the original NUTS algorithm of
 \citet{hoffman_no-u-turn_2014}.  FRUTS discards far fewer points than either
 NUTS or NUTS4 but orients its trajectories randomly, which appears to be a
 significant disadvantage.

\section*{Acknowledgements}

Dr Alexander B. Campbell of Fisheries Queensland (FQ) introduced us to
 Hamiltonian Monte Carlo.  Anonymous reviewers made us aware of unbiased
 simulation, suggested the inclusion of pseudocode, and made many other
 helpful suggestions.  Prof. Jerzy A. Filar of The University of Queensland
 made useful comments on an earlier draft.  Dr Michael F. O’Neill of FQ helped
 with an early (pre-HMC) computational trial.


The authors declare no potential conflict of interest.

\section*{Supplementary material}

\begin{description}

\item[Extra text and figures:] Text and figures not included in body of
 paper. (PDF file)

\item[R scripts:] R scripts for HMC and perfect sampling, with the diabetes
 data set used as an example.  (ZIP file)

\end{description}

\section*{Journal submissions}

We submitted an earlier version of this paper to the journal \textit{Annals of
 Statistics} on 22 May 2020.  It was rejected by the Editor without review but
 with complimentary comments.  The reasons given for rejection were, ``the
 contribution is purely computational'' and ``it lacks appeal to the general
 audience of the AOS and interested readers may also be able to find it more
 easily if it is published elsewhere''.

A similar version was submitted to \textit{Journal of Computational and
 Graphical Statistics} (JCGS) on 8 June 2020.  It was rejected with a
 recommendation for major revision, mainly in layout and style, on 10 March
 2021 (first review period: nine months).

We extensively revised the paper, addressed all of the reviewers' comments,
 and resubmitted it to JCGS on 13 September 2021.  It was again rejected on 6
 May 2022 (second review period: eight months).  The reason given for
 rejection was that it contained a statement that the FRUTS algorithm ran
 ``more efficiently'' than the No U-Turn Sampler (NUTS), which we made on the
 basis that FRUTS had a much lower proportion of discarded trajectory points.
 The reviewer suggested that we either tone down that statement or provide a
 detailed comparison between FRUTS and NUTS.  The Associate Editor requested
 that we provide the detailed comparison.

The first revision included a dedicated section on the recently developed
 technique of ``unbiased simulation'' in which there is much interest, and it
 showed that our methodology could easily accomplish unbiased simulation for
 general continuous problems, thereby greatly expanding the field of
 application of unbiased simulation.  Nevertheless, the reviewer stated, ``The
 question is whether a coupling scheme can be constructed to achieve perfect
 sampling.  If such a scheme cannot be found, I really don't see much point in
 doing this.''

We again extensively revised the paper to include the required comparison and
 address the reviewer's comments.  We comprehensively changed the paper's
 coupling methods in favour of unbiased sampling over the older method of
 coupling from the past (CFTP), included new research showing how to derive
 perfect simulation from unbiased simulation, and presented perfect simulation
 in our results.  We resubmitted to JCGS on 4 September 2022.

The paper was rejected by JCGS for the third and final time on 19 December
 2022 (third review period: three and a half months; total review time with
 JCGS: twenty months; total elapsed time: two years and six months).  The
 reason given for the final rejection was that the reviewer, the Associate
 Editor and the Editor all found the material ``impenetrable''.

\bibliographystyle{chicago}
\bibliography{GeorgeLibrary} 

\newpage

\section*{\centering Supplementary material}

\renewcommand{\theequation}{S\arabic{equation}}
\renewcommand \thesection {S\arabic{section}}
\renewcommand \thefigure {S\arabic{figure}}
\renewcommand \thetable {S\arabic{table}}
\renewcommand \thealgorithm {S\arabic{algorithm}}
\setcounter{equation}{0}
\setcounter{section}{0}
\setcounter{figure}{0}
\setcounter{table}{0}
\setcounter{algorithm}{0}

\section{The full random uniform trajectory sampler}

\subsection{Design}

\begin{algorithm}[ht]
 \caption{FRUTS trajectory for use in Algorithm~1: Setup and forward
  side} \label{algfruts}
 \begin{algorithmic}[1]
  \Procedure{Trajectory}{$q_0, p_0, r$} \Comment{$r$: vector of random
    numbers, indexed by $[j]$}
   \IndentLineComment{Sample the direction vector~$b$ uniformly from the
    surface of the unit sphere.}
   \State $b \gets$ \Call{UnitSphere}{$r[j_\text{dir}]$}
    \Comment{$j_\text{dir}$: Indices of direction-vector random numbers}
   \State $i^+ \gets i^- \gets 0$; $q_\text{traj}[0,] \gets q^+ \gets q^-
    \gets q_0$ \Comment{$q_\text{traj}[i^-:i^+,]$ stores trajectory points.}
   \State $p_\text{traj}[0,] \gets p_0$; $\dot p_0 \gets -DU(q_0)$
    \Comment{$p_\text{traj}[i^-:i^+,]$ stores momentum.}
   \State $p^+ \gets p_0^+ \gets p_0 + \half \delta t\, \dot p_0$;
    $p^- \gets p_0^- \gets p_0 - \half \delta t\, \dot p_0$
    \Comment{$p_0^+, p_0^-$: half time-step offset of $p_0$}
     \label{codetimeoffsets}
   \LineComment{If $b \cdot p_0^+$ and $b \cdot p_0^-$ have the same sign,
    construct two sides to the trajectory.}
   \LineComment{Otherwise, construct the trajectory only on the side with
    the same sign as $b \cdot p_0$.}
   \LineComment{On each constructed side, the leapfrog algorithm runs until
    $b \cdot p^\pm$ changes sign.}
   \If{$\textbf{sgn}(b \cdot p_0^+) \in \{\textbf{sgn}(b \cdot p_0^-),
     \textbf{sgn}(b \cdot p_0)\}$} \Comment{Trajectory has a forward side.}
    \Repeat
     \State $\dot q \gets DK(p^+)$; $q^+ \gets q^+ + \delta t\, \dot q$
      \Comment{Hamiltonian dynamics}
     \State $\dot p \gets -DU(q^+)$; $p^+ \gets p^+ + \delta t\, \dot p$
      \label{codepplus}
     \LineComment{The final trajectory point, half-way along the segment
      over which the sign}
     \LineComment{changes, is included only if it has the original sign of
      $b \cdot p_0^+$.}
     \If{$\textbf{sgn}(b \cdot p^+) = \textbf{sgn}(b \cdot p_0^+)$ \textbf{or}
       $\textbf{sgn}(b \cdot (p^+ - \half \delta t\, \dot p)) =
       \textbf{sgn}(b \cdot p_0^+)$}
      \State $i^+ \gets i^+ + 1$; $q_\text{traj}[i^+,] \gets q^+$;
       $p_\text{traj}[i^+,] \gets p^+ - \half \delta t\, \dot p$
       \Comment{Store point.}
     \EndIf
    \Until{$\textbf{sgn}(b \cdot p^+) \neq \textbf{sgn}(b \cdot p_0^+)$}
     \Comment{Test for change of sign.}
   \EndIf \Comment{\bf{Algorithm continues below}}
  \algstore{frutssto}
 \end{algorithmic}
\end{algorithm}

\begin{algorithm}[ht]
 \ContinuedFloat
 \caption{FRUTS trajectory: Backward side and selection of destination point}
 \begin{algorithmic}[1]
  \algrestore{frutssto}
   \If{$\textbf{sgn}(b \cdot p_0^-) \in \{\textbf{sgn}(b \cdot p_0^+),
     \textbf{sgn}(b \cdot p_0)\}$} \Comment{Trajectory has a backward side.}
    \Repeat
     \State $\dot q \gets DK(p^-)$; $q^- \gets q^- - \delta t\, \dot q$
      \Comment{Hamiltonian dynamics}
     \State $\dot p \gets -DU(q^-)$; $p^- \gets p^- - \delta t\, \dot p$
      \label{codepminus}
     \LineComment{Include the final trajectory point only if it has the
      original sign of $b \cdot p_0^-$.}
     \If{$\textbf{sgn}(b \cdot p^-) = \textbf{sgn}(b \cdot p_0^-)$ \textbf{or}
       $\textbf{sgn}(b \cdot (p^- + \half \delta t\, \dot p))
       = \textbf{sgn}(b \cdot p_0^-)$}
      \State $i^- \gets i^- - 1$; $q_\text{traj}[i^-,] \gets q^-$;
       $p_\text{traj}[i^-,] \gets p^- + \half \delta t\, \dot p$
       \Comment{Store point.}
     \EndIf
    \Until{$\textbf{sgn}(b \cdot p^-) \neq \textbf{sgn}(b \cdot p_0^-)$}
     \Comment{Test for change of sign.}
   \EndIf
   \State $n_\text{pt} \gets i^+ - i^- + 1$ \Comment{Number of points in
    trajectory; note $i^- \le 0$.}
   \LineComment{Select the destination randomly and uniformly from the set
    of trajectory points.}
   \If{$b \cdot q_\text{traj}[i^-,] \le b \cdot q_\text{traj}[i^+,]$}
     \Comment{Order trajectory points by $b \cdot q$; see
      Fig.~1.}
    \State $i_\text{dest} \gets i^- + \textbf{floor}(n_\text{pt} \times
     r[j_\text{sel}])$ \Comment{$r[j_\text{sel}]$: $U(0, 1)$ variate to
     select destination}
   \Else
    \State $i_\text{dest} \gets i^+ - \textbf{floor}(n_\text{pt} \times
     r[j_\text{sel}])$ \Comment{Reverse order compared to above.}
   \EndIf
   \State \textbf{return} $(q_\text{traj}[i_\text{dest},],
    p_\text{traj}[i_\text{dest},])$
  \EndProcedure
 \end{algorithmic}
\end{algorithm}

The FRUTS algorithm (Algorithm~\ref{algfruts}) is conceptually simpler than
 NUTS and better samples the full trajectory, as it includes all the eligible
 points up to the U-turns, rather than discarding up to half of them.  Its
 major innovation is to choose, independently of the trajectory generation, a
 direction vector $b$ for each trajectory, whose dot products with trajectory
 increments must not change sign.  We sample $b$ randomly and uniformly from
 the surface of the $d$-dimensional unit sphere, independently for each
 trajectory.

Limiting the number of points in a trajectory, to bound the storage space and
 execution time in case the time step is too small or the target distribution
 has longer tails than expected, is explained in section~\ref{seclimitpoints}.

Our coding of FRUTS numbers the trajectory points in the order of their dot
 products with the direction vector, not in the order of increasing time as
 stated in section~2.4 and Figure~1 of the paper.  We believed that this
 ordering was more logical given the reliance of the algorithm on the
 direction vector, but it makes no practical difference to the operation of the
 algorithm.

\subsection{Proof of detailed balance}

The general principles by which HMC leapfrog algorithms satisfy detailed
 balance were set out by \citet{duane_hybrid_1987} (see section~2.1).  From
 those principles, \citet[][section 3.1.1]{hoffman_no-u-turn_2014} established
 detailed balance for NUTS by proving that the set, denoted $\mathcal{C}$, of
 discrete points $(q, p)$ in a trajectory is invariant to the choice of
 trajectory origin within $\mathcal{C}$.  In other words, from a discretized
 trajectory $\mathcal{C}$ generated by the algorithm, choose any pair ($q_0$,
 $p_0$) as origin and initial momentum vector of a trajectory, and generate a
 new trajectory $\mathcal{C}'$.  Then an HMC algorithm that selects its
 destination point as $(q, p) \in_R \mathcal{C}$, i.e., randomly and uniformly
 from $\mathcal{C}$, which both NUTS and FRUTS do, satisfies detailed balance
 if $\mathcal{C}' = \mathcal{C}$ for every choice ($q_0$, $p_0$) $\in
 \mathcal{C}$.

We show that FRUTS satisfies the invariance condition $\mathcal{C}' =
 \mathcal{C}$ for detailed balance.

The momentum vectors $p^+$ and $p^-$ at integer-plus-a-half time steps are
 derived from $p_0$ by the derivative $\dot p$ evaluated at integer time
 steps, according to~(2) and lines~\ref{codetimeoffsets}, \ref{codepplus} and
 \ref{codepminus} of Algorithm~\ref{algfruts}.  The reversibility of both HMC
 and the leapfrog algorithm (see section~2.1) proves that
 $p^+$, $p^-$ and all pairs ($q$, $p$) $\in \mathcal{C}$ are reproduced, no
 matter which pair is selected as the origin.  The proof that $\mathcal{C}' =
 \mathcal{C}$ then rests on the FRUTS stopping and starting rules on the
 forward and backward sides of the trajectory, which determine exactly which
 of the discrete trajectory points belong to $\mathcal{C}$ and $\mathcal{C}'$.
 
When $q_0$ is an internal point, not an endpoint of $\mathcal{C}$, the
 reproduction of trajectory points proves that $\mathcal{C}' = \mathcal{C}$
 whatever the choice of internal $q_0$, because the forward and backward
 endpoint stopping rules, determined by the trajectory points, do not depend
 on this choice.  The same argument applies when $q_0$ is an endpoint of
 $\mathcal{C}$ at which $\text{sgn}(b \cdot p_0^+) = \text{sgn}(b \cdot
 p_0^-)$.  In that case the following trajectory point $(q_X, p_X)$ must have
 triggered the stopping rule when $\mathcal{C}$ was constructed, and been
 excluded from $\mathcal{C}$ because $\text{sgn}(b \cdot p_X) \ne \text{sgn}(b
 \cdot p_0^\pm)$.  That point will likewise be excluded from $\mathcal{C}'$.

The remaining case is when $q_0$ is an endpoint of $\mathcal{C}$ and
 $\text{sgn}(b \cdot p_0^+) \ne \text{sgn}(b \cdot p_0^-)$.  Then the FRUTS
 rule for starting a trajectory specifies that $\mathcal{C}'$ is generated
 only on the side with $\text{sgn}(b \cdot p) = \text{sgn}(b \cdot p_0)$.
 Thus $q_0$ is also an endpoint of $\mathcal{C}'$.  Furthermore, by the FRUTS
 stopping rule, if $\mathcal{C}$ originated from a point other than
 $q_0$\thinspace , $q_0 \in \mathcal{C}$ could only have occurred if $q_0$ was
 approached from the side on which $\text{sgn}(b \cdot p_0^\pm) = \text{sgn}(b
 \cdot p_0)$\thinspace ; i.e., the side on which $\mathcal{C}'$ is
 constructed.  Again the reproduction of trajectory points ensures that
 $\mathcal{C}' = \mathcal{C}$.

\subsection{Limiting the number of trajectory sample points}
 \label{seclimitpoints}

This section describes an additional feature of the design of the Full
 Random Uniform Trajectory Sampler (FRUTS), presented in section 3.1
 of the paper.

We impose an initial limit $N$ on the number of points on either side
 of a trajectory, allowing up to $2N + 1$ points over the positive and
 negative sides combined.  Our hope, however, is that the combination
 of the time-step setting and the dot-product conditions of the FRUTS
 algorithm will cause both sides of the trajectory to terminate well
 below this limit.

If the dot-product termination condition is encountered on both sides
 of the trajectory within the initial limit of $N$ points per side,
 the FRUTS algorithm is unaltered.

If the condition is encountered on only one side, we extend the other side
 until it either terminates or would exceed the limit of $2N + 1$ points in
 total, both sides combined.  If it terminates, the FRUTS algorithm is again
 unaltered.  If it doesn't terminate, we sample from the collection of all the
 points on the first side but only the first $N$ points on the second side.
 For each point other than the origin $O$, the probability of selecting that
 point is set at $1/(2N + 1)$ and the balance of the probability is assigned
 to $O$ so that the probabilities sum to~1.  This rule can confer a
 substantial probability (worst case slightly more than 0.5) of staying in the
 same place by choosing the destination $P$ to equal~$O$, but is necessary for
 detailed balance.

If neither side encounters the termination condition, the set of $2N +
 1$ points ($N$ on each side) is used and each point is given equal
 probability $1/(2N + 1)$ of being selected as the destination $P$.

Detailed balance is satisfied because the Markov transition matrix for
 the set of all points on the maximal trajectory from the unadjusted
 FRUTS algorithm is doubly stochastic: all rows and also all columns
 sum to~1.  Therefore its stationary distribution is uniform
 \citep[see, e.g.,][sec. 4.1.1]{pinsky_introduction_2011}, which is
 the condition for detailed balance.

\begin{figure}[ht]
 \centering
 \includegraphics[width=\textwidth, trim=1.5cm 7.5cm 2cm 1.5cm,
  clip=true]{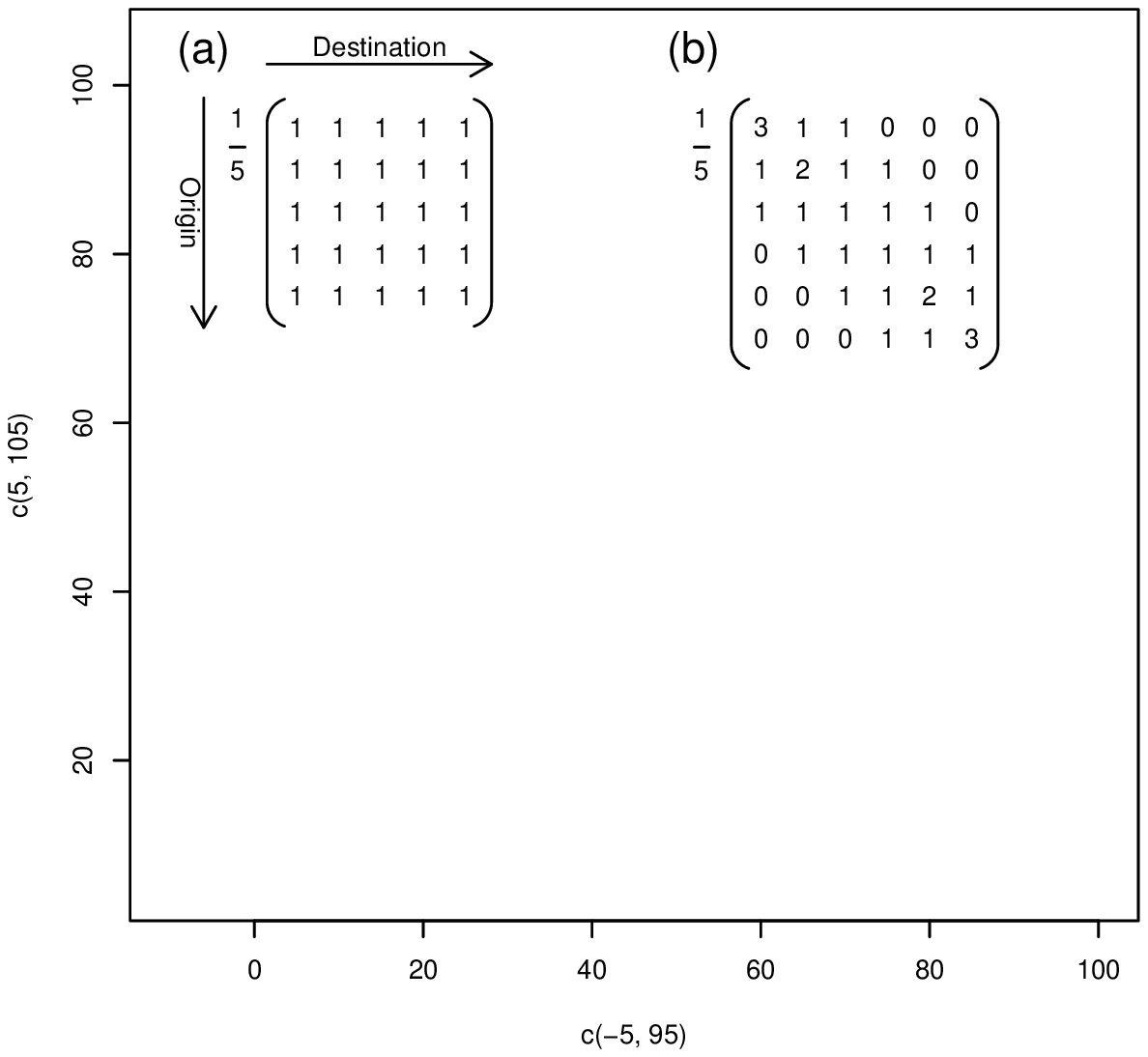}
 \caption[Markov transition matrices with limited number of points]
 {Markov transition matrices showing implementation of an artificial
  limit on the number of trajectory points, to limit storage space and
  execution time in case the time step has been chosen too small or
  the target distribution is unexpectedly long-tailed.  Here the limit
  is $2N + 1$ = 5 points, and the whole trajectory consists of either
  (a) 5 or (b) 6 points.  In (a), the limit fits within the full
  length of the trajectory, and the destination is selected from a
  discrete uniform distribution.  In (b), the trajectory contains more
  points than the limit: then a maximum of $N = 2$ points on each side
  of the origin is allowed.  Both matrices are doubly stochastic, so
  have uniform stationary distribution and thereby satisfy detailed
  balance.} \label{figtransition}
\end{figure}

The doubly stochastic nature of the transition matrices and the change
 in formulation when the trajectory is longer than $2N + 1$ points are
 illustrated in Figure~\ref{figtransition}.

\newpage
\section{Setting the time step}\label{supsectimestep}

For convenience, we reproduce two equations from the main paper here:
 the kinetic energy equation
\begin{equation}
 K = |p|^\beta/\beta \label{supeqkpower}
\end{equation}
and the velocity equation
\begin{equation}
 \dot q = \left(dK/{d|p|}\right) p/|p| = |p|^{\beta - 2} p.
  \label{supeqqdotkpower}
\end{equation}

We assume that the location coordinates $q$ are scaled to make the
 variance of $q$ approximately equal to the identity matrix; e.g., by
 first optimizing to find the maximum-likelihood point, then scaling
 by the square-root of the Hessian matrix at that point.  Such a
 procedure is used in, e.g., the software AD Model
 Builder~\citep{fournier_ad_2011}.

Similarly to equation (\ref{supeqkpower}) for kinetic energy, for the
 purpose of setting the time step we approximate the potential energy
 (negative log-likelihood) by the form
\begin{equation}
 U = |q^\alpha| / \alpha, \quad |q| = (\alpha U)^{1 / \alpha}.\label{equpower}
\end{equation}
\noindent The case $\alpha = 2$ corresponds to a normal distribution.
 Values $\alpha < 2$ can be used if the target distribution is
 long-tailed (e.g., our $t$-distribution example).  The
 form~(\ref{equpower}) calibrates the negative log-likelihood to be
 zero at the maximum-likelihood point.

The boundary of the region in $q$-space that a trajectory with
 Hamiltonian value $H$ can traverse is hit when $U = H$ and $K = 0$;
 i.e., maximum potential and minimum kinetic energy.  This region has
 diameter, from~(\ref{equpower}),
\begin{equation}
D = 2 (\alpha H)^{1 / \alpha}.\label{eqdiameter}
\end{equation}

The speed of the HMC particle on its trajectory is, by
 (\ref{supeqqdotkpower}),
\begin{equation}
 |\dot q| = |p|^{\beta - 1} = (\beta K)^{1 - 1/\beta} =
 (\beta H)^{1 - 1/\beta} (K/H)^{1 - 1/\beta}.\label{eqqdotkh}
\end{equation}

By (\ref{supeqkpower}) and~(\ref{equpower}), incorporating the surface
 area of a $d$-dimensional sphere, the densities of $|p|$ and $|q|$
 are proportional to $|p|^{d - 1} \exp\big(-|p|^\beta / \beta \big)$
 and $|q|^{d - 1} \exp\big(-|q|^\alpha / \alpha \big)$ respectively.
 Transforming back to the variables $K$ and $U$, the density of $K$ is
 proportional to ($K > 0$)
\begin{equation*}
 |p|^{d - 1} \exp{(-|p|^\beta/\beta)} \big/ {(dK/d|p|)} =
 |p|^{d - \beta} \exp(-|p|^\beta / \beta) \propto
 K^{d/\beta - 1}\exp(-K),
\end{equation*}
\noindent a gamma distribution with shape parameter $d/\beta$ and rate
 parameter 1.  Similarly, $U$ has an independent gamma distribution
 with shape parameter $d/\alpha$ and rate parameter~1.

Returning to~(\ref{eqqdotkh}) and using properties of the gamma and
 beta distributions \citep[see,
 e.g.,][]{encyclopedia_of_mathematics_gamma-distribution_2020}, the
 distributions of $H = U + K$ and $K/H$ are, independently, gamma with
 shape $d/\beta + d/\alpha$ and beta with parameters ($d/\beta$,
 $d/\alpha$).  The expectation of $|\dot q|$ conditional on $H$ is
\begin{align*}
 E\big(|\dot q|\, \big|\, H\big) &= (\beta H)^{1 - 1/\beta}\,
  \frac{\Gamma(d/\beta + d/\alpha)}
  {\Gamma(d/\beta)\, \Gamma(d/\alpha)}
  \int_0^1 x^{1 - 1/\beta} x^{d/\beta - 1} (1 - x)^{d/\alpha - 1}\,dx \\
 &= (\beta H)^{1 - 1/\beta}\, \frac{\Gamma(d/\beta + 1 - 1/\beta)\,
  \Gamma(d/\beta + d/\alpha)} {\Gamma(d/\beta)\, \Gamma(d/\beta + d/\alpha +
  1 - 1/\beta)}\,.
\end{align*}

Let $h$ be the reciprocal of the number of points desired in a
 trajectory.  Then, still conditional on $H$ and
 using~(\ref{eqdiameter}), the time step $\delta t$ can be set to
\begin{equation}
 \delta t = hD / {E\left(|\dot q|\, \big|\, H\right)} = 2h (\alpha H)^{1/\alpha}
 (\beta H)^{1/\beta - 1} \frac{\Gamma(d/\beta)\, \Gamma(d/\beta + d/\alpha + 1 -
 1/\beta)} {\Gamma(d/\beta + 1 - 1/\beta)\, \Gamma(d/\beta + d/\alpha)} \,.
 \label{tconditional}
\end{equation}

If $\alpha = \beta = 2$, this does not depend on $H$ and it can be
 expected that the number of trajectory points will vary little
 between trajectories.  For general $\alpha$ and $\beta$, which may be
 needed for long-tailed distributions, we use (\ref{tconditional})
 with the gamma distribution of $H$ to find
\begin{equation*}
 E(1/\delta t) = \frac{\beta^{1 - 1/\beta} \alpha^{-1/\alpha}
 \Gamma(d/\beta + 1 - 1/\beta)} {2h \Gamma(d/\beta)\, \Gamma(d/\beta +
 d/\alpha + 1 - 1/\beta)} \int_0^\infty H^{1 - 1/\beta - 1/\alpha}
 H^{d/\beta + d/\alpha - 1} e^{-H}\,dH. \label{eqerecipdeltat}
\end{equation*}
\noindent The integral is another gamma function and taking
 reciprocals yields the final time step:
\begin{equation}
 \delta t = 2h\beta^{1/\beta - 1} \alpha^{1/\alpha}\, \frac{\Gamma(d/\beta)}
 {\Gamma\big(\{d - 1\} / \beta + 1\big)} \cdot
 \frac{\Gamma\big(\{d - 1\}/\beta + d/\alpha + 1\big)}
 {\Gamma\big(\{d - 1\}/\beta + \{d - 1\}/\alpha + 1\big)}\,.
 \label{supeqdeltatfinal}
\end{equation}

\newpage
\section{CFTP and ROCFTP algorithms}\label{secunbiasedalg}

\begin{algorithm}[!ht]
 \caption{Coupling from the past, calling HMC with rounding (Algorithm~3)}
  \label{algcftp}
 \begin{algorithmic}[1]
  \Procedure{CFTP}{$n_\text{st}, Q_0, n_\text{it}, N_\text{it}, R,
    r_\text{ro}, w$}
   \Comment{$n_\text{st}, Q_0$: number and matrix of starting}
   \LineComment{points; $n_\text{it}, N_\text{it}$: number and maximum allowed
    number of iterations (trajectories)}
   \State $Q \gets Q_0$; $n_\text{coal} \gets 1$
    \Comment{Matrix to hold results; number of coalesced results}
   \For{$i_\text{st} \gets 1:n_\text{st}$} \Comment{Loop over
     different starting points.} \label{cftpistloop}
    \IndentLineComment{Use same random numbers $R$ for each starting point;
     always end at row $N_\text{it}$.}
    \State $Q[i_\text{st},] \gets$ \Call{HmcRound}{$Q[i_\text{st},],
     n_\text{it}, R[(N_\text{it} - n_\text{it} + 1):N_\text{it},],
     r_\text{ro}, w$}
     \label{cftphmccall}
    \If{$i_\text{st} \ge 2$ \textbf{and} $Q[i_\text{st},] = Q[i_\text{st} -
      1,]$} \Comment{Count coalesced outcomes.}
     \State $n_\text{coal} \gets n_\text{coal} + 1$ \Comment{Increment when
      same as from previous starting point.}
    \EndIf
   \EndFor \label{cftpistloopend}
   \State \textbf{return} $(Q, n_\text{coal})$
    \Comment{Increase $n_\text{it}$ and rerun until all coalesced
    ($n_\text{coal} = n_\text{st}$).}
  \EndProcedure
 \end{algorithmic}
\end{algorithm}

\begin{algorithm}[!hp]
 \caption{Read-once coupling from the past (ROCFTP) (adjusted
  Algorithm~\ref{algcftp})} \label{algrocftp}
 \begin{algorithmic}[1]
  \Procedure{ROCFTP}{$n_\text{st}, Q_0, n_\text{it}, N_S, d_R, w$}
    \Comment{$n_\text{st}, Q_0$: number and matrix of starting}
   \LineComment{points; $n_\text{it}$: block length (number of trajectories);
    $N_S$: required sample size;}
   \LineComment{$d_R$: number of random numbers per trajectory; $w$: rounding
    width}
   \State $Q \gets \textbf{zeros}(n_\text{st} + 1, d)$; $Q_S \gets
    \textbf{zeros}(N_S, d)$ \Comment{Storage for current and final results}
   \State $Q[n_\text{st} + 1,] = Q_0[n_\text{st},]$ \Comment{Initialize the
    final-result row; don't reset during blocks.}
   \State $i_S = 0$ \Comment{Set final-result counter; need one coalesced
    block before we store any.}
   \Repeat
    \State $Q_S[\max(i_S, 1),] = Q[n_\text{st} + 1,]$ \Comment{Store initial
     value in case this block coalesces.}
    \State $R \gets \textbf{rand}(n_\text{it}, d_R)$ \Comment{Generate random
     numbers for HMC, dimensions $n_\text{it} \times d_R$.}
    \State $r_\text{ro} \gets \textbf{rand}(d + 1)$ \Comment{Generate random
     numbers for rounding, dimension $d + 1$.}
    \State $Q[1:n_\text{st},] \gets Q_0$; $n_\text{coal} \gets 1$
     \Comment{Initialize block and number of coalesced results.}
    \For{$i_\text{st} \gets 1:n_\text{st} + 1$} \Comment{Loop over different
      starting points.}
     \State $Q[i_\text{st},] \gets$ \Call{HmcRound}{$Q[i_\text{st},],
      n_\text{it}, R, r_\text{ro}, w$}
     \If{$i_\text{st} \ge 2$ \textbf{and} $i_\text{st} \le n_\text{st}$
       \textbf{and} $Q[i_\text{st},] = Q[i_\text{st} - 1,]$}
      \State $n_\text{coal} \gets n_\text{coal} + 1$ \Comment{Count coalesced
       outcomes (where same as previous).}
     \EndIf
    \EndFor
    \If{$n_\text{coal} = n_\text{st}$} \Comment{Test coalescence (same results
      from all starting points)}.
     \State $i_S = i_S + 1$ \Comment{One final result per coalesced block,
      after the first one}
    \EndIf
   \Until{$i_S > N_S$}
   \State \textbf{return} $Q_S$
  \EndProcedure
 \end{algorithmic}
\end{algorithm}

\newpage
\section{Results for NUTS and Raw HMC}
 \label{secresultsnutsraw}

\begin{table}[hb]
 \caption[Results for NUTS and Raw HMC]
 {Perfect simulation results for standard distributions with the NUTS and Raw
  HMC algorithms.  Symbol $d$ is number of dimensions, $\rho$ is pairwise
  correlation coefficient (aspect ratio of variance ellipsoid in parentheses),
  $\nu$ is degrees of freedom, $\alpha$ is the distribution-tail parameter
  (noted when $\alpha \ne 2$), and $\mu$ is the nonzero mode.  Perfect-sample
  size is listed (number of independent sample sets in parentheses, k denotes
  1000), then numbers of likelihood-derivative evaluations per perfect sample
  point for the NUTS and Raw HMC algorithms (dash indicates not conducted),
  and the maximum number of blocks for coalescence of chains in Algorithm~4 of
  the paper (bold indicates problem cases).}
  \label{tabresultsnutsraw}
 \centering
 \begin{tabular}{l r l r r r r r}
 \hline
 Name & $d$ & Parameters set & Sample & NUTS & Raw & Max \\
 \hline
 Standard normal & 1 & -- & 140k (10k) & 914 & 300 & 3 \\
  & 10 & -- & 14k (1k) & 1350 & 521 & 3 \\
  & 100 & -- & 14k (1k) & 1439 & 642 & 2 \\
 Correlated normal& 2 & $\rho = 0.6$ (4) & 14k (1k) & 1738 & 652 & 3 \\
  &  & $\rho = 0.95$ (39) & 14k (1k) & -- & 841 & 3 \\
  & 100 & $\rho = 0.45$ (82.82) & 280 (20) & 42222 & -- & 2 \\
 $t$-distribution & 1 & $\nu = 4$ & 14k (1k) & 1118 & 1092 & \textbf{12} \\
  & 10 & $\nu = 4$, $\alpha = 1.5$ & 14k (1k) & 4756 & 5047 & \textbf{7} \\
  & 100 & $\nu = 4$, $\alpha = 1.25$ & 14k (1k) & 32361 & -- & 5 \\
 Normal mixture & 1 & $\mu = 4$ & 14k (1k) & -- & 1222 & 4 \\
  & & $\mu = 6$ & 2.8k (200) & 6396 & 9550 & \textbf{5} \\
  & 10 & $\mu = 6$ & 1.4k (100) & -- & 12046 & \textbf{6} \\
  & 100 & $\mu = 6$ & 280 (20) & 13908 & -- & 4 \\
 \hline
 \end{tabular}
\end{table}

\section{Histograms of the residual sum of squares in the Bayesian Lasso}
 \label{sechistplot}

This section presents and describes the histograms of the residual sum of
 squares, mentioned in section~5 of the paper.

\begin{figure}[p]
 \centering
 \includegraphics[width = 13cm, 
  trim = 0cm 0.3cm 0cm 0.6cm]{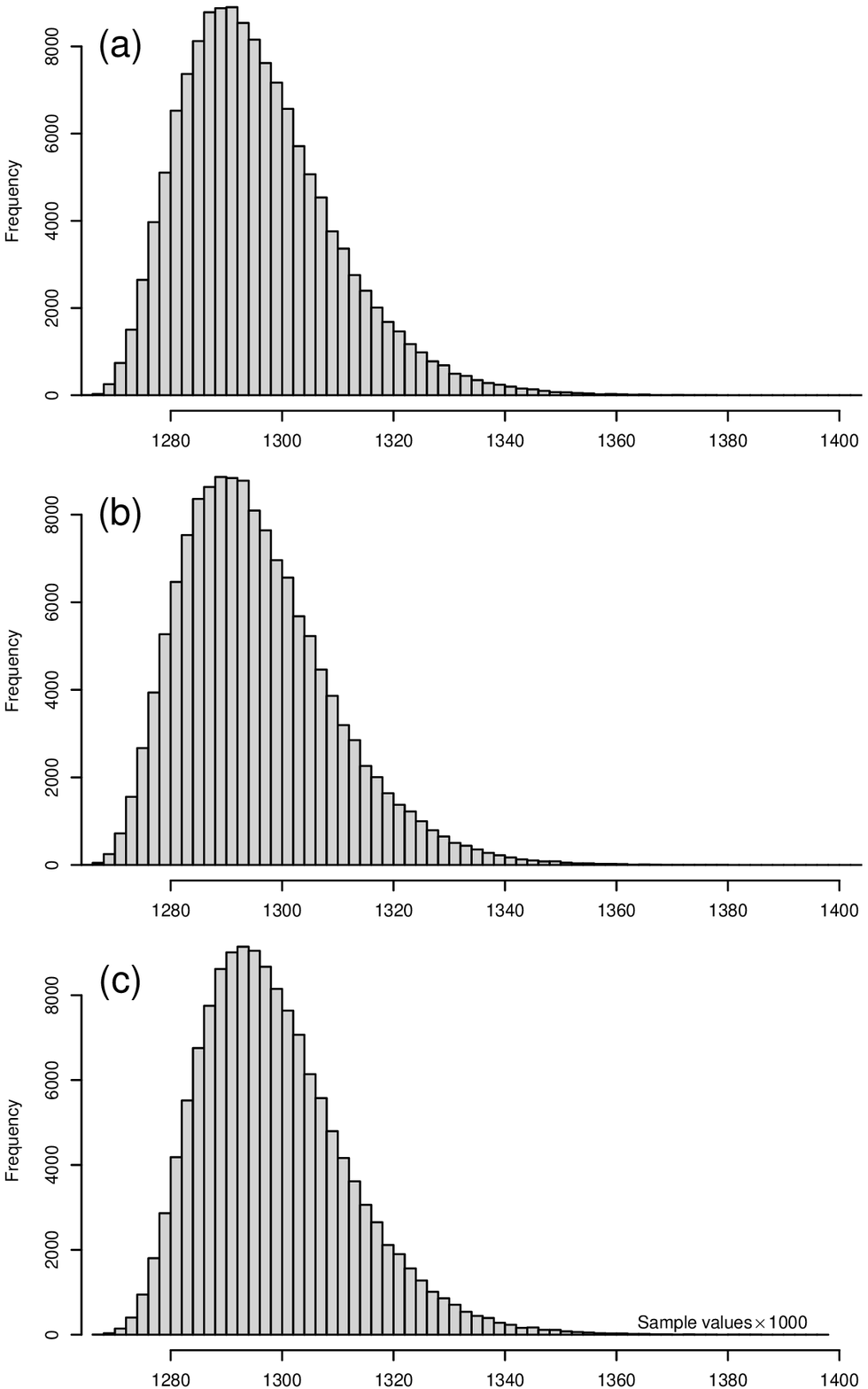}
 \caption[Histograms of the Bayesian Lasso residual sum of squares]
 {Perfect-sample histograms of the residual sum of squares for the Bayesian
  Lasso, $S$, for different settings of the Lasso parameter $\lambda$:
  (a)~$\lambda = 0$, (b)~$\lambda = 0.237$, (c)~$\lambda =
  5$.}  \label{fighistlassos}
\end{figure}

Histograms of the residual sum of squares, $S$, are shown in
 Figure~\ref{fighistlassos}, for $\lambda = 0$ (no Lasso term), 0.237 and 5.

The distributions of~$S$ for $\lambda = 0$ and $\lambda = 0.237$ are
 practically indistinguishable.  The one for $\lambda = 5$ shows a slight
 shift towards larger values of~$S$; i.e., slightly worse fit of the linear
 regression.  The mean values of~$S / 1000$ are 1295.65, 1295.52 and~1298.76
 for $\lambda = 0$, 0.237 and~5 respectively.

\end{document}